\documentclass[11pt,a4paper]{article}
\usepackage{cite}
\usepackage{amsmath}
\usepackage{amssymb}
\usepackage{mathtools}
\usepackage{stmaryrd}
\usepackage{float}
\usepackage{hyperref}
\usepackage{placeins}
\usepackage{enumitem}
\usepackage{array}

\title{Diffusion Schr\"odinger Bridges with enhanced posterior sampling for metasurface inverse design}
\author{
	\textbf{Mathys Le Grand$^{1,2,*}$, \quad Pascal Urard$^{2}$,\quad Denis Rideau$^2$,}\\
	\textbf{\quad Loumi Tr\'emas$^2$, \quad Damien Maitre$^2$, \quad  Adam Fuchs$^2$,}\\
	\textbf{Louis-Henri Fernandez-Mouron$^2$,}\quad \textbf{R\'egis Orobtchouk$^1$} \\
	$^1$Institut des nanotechnologies de Lyon \quad $^2$STMicroelectronics \\
	\texttt{mathys.le-grand@insa-lyon.fr}}
\date{}

\begin{document}
	\maketitle
	\begin{abstract}
		
		Metasurface inverse design is challenged by the intricate relationship between structural parameters and electromagnetic responses, as well as the high dimensionality of the optimization space. Local models, while commonly employed, quickly become infeasible for complex and locally coupled structures. Conventional iterative optimization techniques, on the other hand, are computationally intensive, time-consuming, and susceptible to convergence in local minima.
		
This study explores a versatile generative methodology based on enhanced posterior sampling within the Schr\"odinger Bridge framework. By decomposing posterior sampling into amplitude and directional contributions, we effectively integrated different kind of posterior sampling. This approach is further supported by refined training strategies to enhance performance and reduce the complexity of hyperparameter optimization.

The proposed framework demonstrates exceptional accuracy and robustness, representing a significant advancement in metasurface design. Notably, it enables high-precision inverse design for large-scale configurations of up to $350 \times 350$ pillar arrays, despite being trained on significantly smaller arrays of $23 \times 23$ pillars.

	\end{abstract}
	\section{Introduction}\label{sec:intro}
	
	Metasurfaces are ultrathin, two-dimensional arrangements of engineered subwavelength structures often called meta-atoms that manipulate electromagnetic waves in ways not achievable with natural materials. By precisely tailoring the geometry, size, and arrangement of these meta-atoms, metasurfaces can control the amplitude, phase, polarization, and direction of light at the interface, enabling unprecedented control over wavefronts \cite{hu2021review,chen2016review}.
	
	Unlike their three-dimensional metamaterial counterparts, metasurfaces offer significant advantages in fabrication simplicity and reduced losses due to their planar nature and minimal thickness. This makes them particularly attractive for applications in the optical regime where fabrication constraints and material absorption are critical challenges.
	
	The ability of metasurfaces to impose abrupt changes on electromagnetic waves has led to breakthroughs in diverse areas such as flat lenses (metalenses)\cite{lepers2023metasurface}, holography \cite{zheng2015metasurface,guo2020flexible}, beam steering \cite{rideau2024approaches}, polarization control \cite{yang2014dielectric,zhu2013linear}. Their compact form factor and versatile functionality position metasurfaces as a transformative technology for next-generation photonic devices and systems.

Inverse design seeks physical configurations that achieve specified optical responses. While global optimization methods such as Bayesian optimization \cite{jones1998efficient} are effective, their applicability is constrained by the computational demands of conventional simulation techniques like Finite Difference Time Domain (FDTD) \cite{gedney2011introduction}. To address this limitation, deep learning-based surrogate models have been developed to accelerate simulations \cite{rideau2024approaches} and extend Bayesian optimization capabilities \cite{Damienrouter}. However, Bayesian optimization remains practical only for problems with relatively low dimensionality, typically up to around 60 parameters \cite{santoni2024comparison}. The metasurfaces studied here involve hundreds to millions of parameters, rendering Bayesian optimization computationally infeasible.

Alternatively, direct structure prediction using deep learning models has been investigated but encounters challenges due to the inherently one-to-many nature of the inverse problem, often leading to unstable or non-convergent training \cite{liu2018training}.
	To overcome these limitations, generative methodologies including Variational Autoencoders (VAEs) \cite{ma2019probabilistic}, Generative Adversarial Networks (GANs) \cite{so2019designing}, and Diffusion Models (DMs) \cite{grand2026enhanced} have been explored. Recently, Schr\"odinger Bridges (SBs), a generalization of DMs, have gained prominence in generative modeling \cite{chen2021likelihood}, yielding Diffusion Schr\"odinger Bridges (DSBs) \cite{de2021diffusion}. This framework has been further adapted to paired data contexts \cite{liu20232} by extending the DM approach.
	
	DSB frameworks provide a probabilistic approach to modeling complex distributions by solving stochastic optimal transport problems, offering a principled alternative to diffusion-based generative models. Unlike conditional diffusion models, which operate on Gaussian distributions, DSB methods explicitly incorporate the structure of the input data. Moreover, similar to diffusion models DMs, DSBs frameworks support posterior sampling \cite{chung2024direct}, allowing the extension of DM posterior sampling techniques to improve DSB performance.
	
	DSB methods without posterior sampling have demonstrated superior sample quality in inverse problems such as medical imaging reconstruction \cite{mirza2023learning}, physical field reconstruction from sparse measurements \cite{li2025physics}, and image deconvolution for correcting detector effects in astrophysics \cite{diefenbacher2024improving}. However, the application of DSB methods with posterior sampling to inverse design problems remains unexplored.
	
	This paper investigates the application of DSB methods to the inverse design of metasurfaces. We incorporate a consistency term into the loss function \cite{grand2026enhanced} and develop enhanced posterior sampling strategies to ensure adherence to input conditions. Our study initially validates the performance gains of DSB methods with posterior sampling on small metasurfaces, followed by the introduction of a scalable approach for the inverse design of larger metasurfaces.
	
\begin{figure}[]
		\centering
		\includegraphics[scale=0.45]{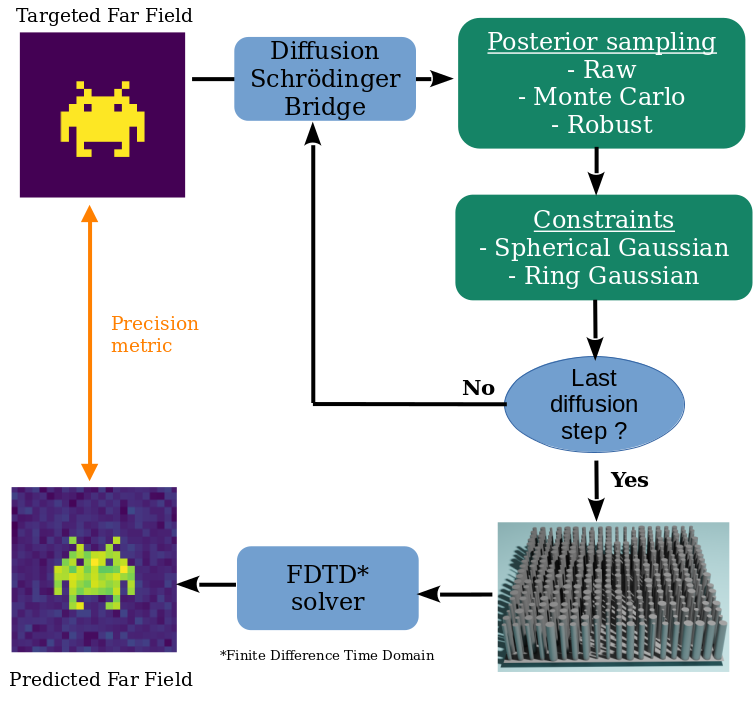}
		\caption{Validation workflow for the inverse design pipeline employing DSBs.}
		\label{fig:verif_process}
	\end{figure}

 These methodologies are applied to the inverse design of a beam-shaping metasurface, with the objective of attaining a predefined far field light amplitude distribution. The metasurface architecture comprises dielectric pillars of adjustable radii, illuminated by a monochromatic plane wave at normal incidence, characterized by a wavelength $\lambda$ and an inter-pillar spacing of $\lambda/2$. The optimized parameters are validated by comparing the simulated far field response to the target distribution, utilizing the $R^2$ metric. Given the nonlinear nature of the problem, the $R^2$ metric ranges over $(-\infty, 1]$, where values closer to 1 signify better alignment with the target. The evaluation process is depicted in Figure~\ref{fig:verif_process}. The DSBs with enhanced posterior sampling, as presented in this study, will be benchmarked against reference methods detailed in Appendix~\ref{App:ref}.

	\section{Contributions}
The primary objective of this paper is to extend the diffusion model framework introduced in~\cite{grand2026enhanced} by adapting analogous methodologies to DSBs. Additionally, we introduce one novel posterior sampling technique (Robust Posterior Sampling) and two posterior sampling constraints (Disk and Ring Gaussian constraints), specifically designed for combined application.

The main contributions of this work, denoted as (ours) in Table~\ref{tab:related}, are compared with the findings presented in~\cite{grand2026enhanced}. For a broader comparison with related research on inverse design using diffusion models, refer to the Related Works section in~\cite{grand2026enhanced}. 
\textbf{The code for this study is open-sourced at \cite{le_grand_2026_18504073}.}
	
		\begin{table}[h] 
		\centering
		
	\begin{tabular}{|p{3cm}|p{4cm}|p{4cm}|} 
		\hline
		Topic  & \cite{grand2026enhanced} & This work \\ \hline
		Figure of Merit &Far Field power distribution for beam shaping& Far Field power distribution for beam shaping  \\ 
		\hline
		Structure & Pillars & Pillars \\ 
		\hline
		Method & Diffusion Models & Diffusion Schr\"odinger Bridge \\
		\hline
		Posterior Sampling & \begin{minipage}[t]{4cm} 
			\begin{itemize}[leftmargin=2mm, itemsep=0pt, parsep=0pt, topsep=2pt, partopsep=0pt]
				\item Raw \cite{chung2022diffusion}
				\item Monte Carlo \cite{song2023loss}
			\end{itemize}
			\vspace{0.2mm}
		\end{minipage} &
		\begin{minipage}[t]{4cm} 
			\begin{itemize}[leftmargin=2mm, itemsep=0pt, parsep=0pt, topsep=2pt, partopsep=0pt]
				\item Raw 
				\item Monte Carlo
				\item Robust (ours)
			\end{itemize}
			
		\end{minipage} \\
		 \hline
		 Posterior Sampling & \begin{minipage}[t]{4cm} 
		 	\begin{itemize}[leftmargin=2mm, itemsep=0pt, parsep=0pt, topsep=2pt, partopsep=0pt]
		 		\item Spherical Gaussian constraint \cite{yang2024guidance}
		 	\end{itemize}
		 	\vspace{0.2mm}
		 \end{minipage} &
		 \begin{minipage}[t]{4cm} 
		 	\begin{itemize}[leftmargin=2mm, itemsep=0pt, parsep=0pt, topsep=2pt, partopsep=0pt]
		 		\item Spherical Gaussian constraint
		 		\item Disk Gaussian costraint (ours)
		 		\item Ring Gaussian constraint (ours)
		 	
		 	\end{itemize}
		 	
		 \end{minipage} \\
		 	\hline
		 	Robustness study & \textbackslash& \checkmark \\ 
		 \hline
		 Scaling (Higher Degrees of Freedom) & High precision up to $98 \times 98$ pillars & High precision up to  $350 \times 350$ pillars \\ 
		 \hline
		 
	\end{tabular}
	\caption{Comparative overview of the key topics addressed in related research on metasurface inverse design using diffusion models with enhanced posterior sampling.}
	\label{tab:related}
		\end{table}

\section{Diffusion Schr\"odinger Bridge}
The Schr\"odinger Bridge problem, first introduced in \cite{schrodinger1932theorie} and later surveyed in \cite{leonard2013survey}, is an entropy-regularized stochastic optimal transport problem. It aims to identify the most likely stochastic evolution connecting an initial and a target distribution through a diffusion process, thus bridging the two distributions over time. Early solutions to the SB problem employed methods such as Iterative Proportional Fitting \cite{chen2021optimal}. More recently, diffusion model-based approaches, known as DSBs, were introduced in \cite{de2021diffusion}, enabling generative processes. However, to apply diffusion bridges to inverse design problems, they must operate on paired data. This limitation was addressed in \cite{liu20232} and further refined using posterior sampling techniques in \cite{chung2024direct}.

This approach extends diffusion models by incorporating data structure and constraints directly into the generative process, eliminating the need for conditioning the score at each step with external inputs, as required in conditional diffusion models \cite{grand2026enhanced}. By embedding data structure and exploiting nonlinear diffusion dynamics, as demonstrated in Section~\ref{sec:steps_study}, DSBs outperform conventional diffusion models while using fewer generation steps. In this work, the score network is conditioned on the far field amplitude. A detailed mathematical treatment of DSBs is provided in Appendix~\ref{App:maths}.

	\subsection{Score Conditional Diffusion Schr\"odinger Bridge} \label{sec:condition}
	For inverse design, the generation process must be biased to satisfy a given condition $c$. In DMs, this is typically achieved by providing the condition $c$ as an additional input to the score network, $ \epsilon_\theta(x_t,c, t)$. 
	
	In contrast, Diffusion Schr\"odinger Bridge incorporate the condition directly into the input, setting $x_T^{\mathrm{DSB}} = c$, whereas in standard DMs, the initial sample $x_T^{\mathrm{DM}}$ is drawn from a Gaussian distribution $\mathcal{N}({0}, {I})$.
	Thus, DSB receive the condition only once at the initial sampling step, whereas DMs incorporate the condition at every step through the score prediction. Although the conditioning methods differ between DMs and DSBs, the score network of DSB can be modified to also accept the condition $c$ in the score prediction. This enables DSB to leverage the condition both through the input and the score network.
	
\begin{figure}[H]
		\centering
		\includegraphics[scale=0.45]{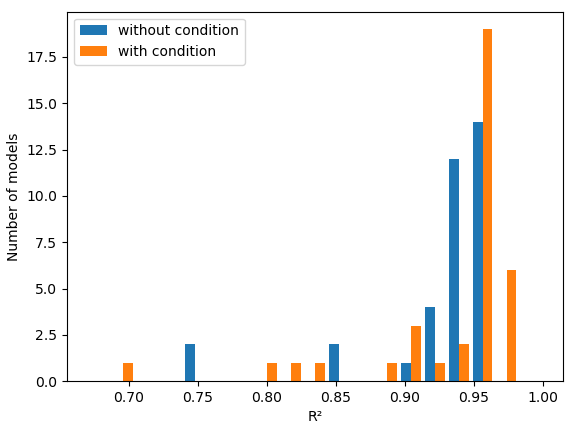}
		\caption{Final performance metrics are compared with and without score conditioning. Models employing conditional scores demonstrate superior results.}
		\label{fig:score_condition}
	\end{figure}

	The score prediction conditioning for DSB  has a clear impact on performance general distribution since as shown on Figure \ref{fig:score_condition}, there are as many poor model performer of DSB with or without score conditioning. However score conditioning increases the performance of the already best performing models as the first twelve out of 100 models have score conditioning. Hence giving the condition as input and then at every sampling steps helps enhancing performance.
	
	\section{Posterior sampling}
	
	The results presented in this section are based on the best-performing model selected via the training improvement techniques presented in \cite{grand2026enhanced} and derived for DSBs in Appendix \ref{App:training}. Model performance remained consistent across the three enhanced sampling methods introduced. 
	
	As discussed in the training improvements section, these advanced sampling methods were previously evaluated for metasurface inverse design using DMs in \cite{grand2026enhanced}. Here, we assess the applicability and effectiveness of these posterior sampling techniques on DSBs.
	
Equation~\eqref{eq:predicted_mean_posterior_sampling_q} illustrates the modification of the predicted mean through posterior sampling guidance. Additional details and results are provided in Appendix~\ref{App:posterior_sampling}.
	\begin{equation}
		\label{eq:predicted_mean_posterior_sampling_q}
		\mu_\theta^{ps}(x_t, c, t) = \mu_\theta(x_t, c, t) - q_t \nabla_{x_t} \| c - S_\phi(x_{0|t}) \|^2_2,
	\end{equation}

In this work, $S_\phi$ represents a surrogate model tasked with predicting the far field from metasurface parameters, employing the same architecture as detailed in \cite{grand2026enhanced}. The Score Conditional DSB utilizes an identical network architecture to that of the DMs presented in \cite{grand2026enhanced}. For the DSBs, the same architecture is retained, albeit with the conditional component omitted.

Prior works \cite{chung2022diffusion,song2023loss} demonstrate that, while using $q_t = 1$, stability and performance improve when employing a weighting scheme defined as 
$q_t = \frac{\mathrm{const}}{\|c - S_\phi(x_{0|t})\|_2},$
referred to here as normalized guidance.

	However, as shown in Table ~\ref{table::PS}, normalized guidance slightly decreases precision on the final metric for $23 \times 23$ pillar metasurfaces. Consequently, posterior sampling without normalized guidance, which performs better in this setting, is compared to other methods in Table \ref{table}. These findings for DSBs align with observations for diffusion models reported in \cite{grand2026enhanced}.
	
	Posterior sampling demonstrates promising results. In contrast, ancestral sampling \cite{ho2020denoising} fails to generate shapes closely matching the conditioning input.  In the subsequent subsections, we introduce two distinct approaches to refine posterior sampling: modifying the guidance direction and adjusting its amplitude. We further explore how these strategies can be synergistically leveraged.
	
	\begin{table}[h]
		\centering
		\begin{tabular}{|p{4cm}|l|l|}
			\hline
			Guidance & Normalized & Unormalized \\
			\hline
			$mean(R^2)$  &0.897 & 0.951  \\ 
			\hline
		\end{tabular}
		\caption{Reported $R^2$ values are derived from FDTD simulations conducted after inverse design using DSB PS.}
		\label{table::PS}
	\end{table}
	
	\subsection{Enhanced Guidance with Robust posterior sampling}
		
Building on the Monte Carlo posterior sampling (PS MC) method of \cite{song2023loss}, which samples around $x_{0|t}$ at each step to estimate guidance, we propose an improved aggregation strategy for sample contributions. Contrary to exponential averaging that amplifies the influence of samples with higher loss, our approach attenuates their impact. Specifically, we compute a weighted average of sample losses, where weights decrease with the distance between each sample $x_{0|t}^i$ and the Gaussian mean $x_{0|t}$. This emphasizes samples closer to the mean, fostering the identification of robust solutions within the posterior sampling framework. 

A robust solution is defined as an optimal input whose output remains stable under small input perturbations. This concept is particularly critical in metasurface inverse design, where simulations do not capture fabrication variabilities. Designing metasurface structures that maintain performance despite such uncertainties is therefore essential. Accordingly, the predicted mean $\mu_\theta$ is modified as

\begin{equation}
	\mu_\theta^{stable}(x_t, c, t) = \mu_\theta(x_t, c, t) \\
	+ \nabla_{x_t} \left( Loss_{central} + s(t)Loss_{robust} \right),
\end{equation}
with :
\begin{subequations}
	\begin{align}
		 Loss_{central} = \| c - S_\phi(x_{0|t}) \|^2_2\\
		 Loss_{robust} = \frac{\sum_{i=1}^{N}\left(\|x_{0|t} - x_{0|t}^{i}\|_2\| c - S_\phi(x_{0|t}^{i}) \|^2_2 \right)}{\sum_{i=1}^{N}\|x_{0|t} - x_{0|t}^{i}\|_2}
	\end{align}
\end{subequations}

Robust Posterior Sampling, analogous to Monte Carlo posterior sampling, generates samples according to
$x_{0|t}^i \sim \mathcal{N}\big(x_{0|t}, g_t^2 \mathbf{I}\big),$
where $g_t$ represents an optimizable schedule. This method further incorporates a robustness schedule $s(t)$, which dynamically adjusts the emphasis on robustness throughout the generation process. Following the optimization of both $g_t$ and $s(t)$, Robust Posterior Sampling with unnormalized guidance achieves the highest performance on the $23 \times 23$ metasurface design problem, as detailed in Table~\ref{table::PS_robust}. Similar to PS MC$_N$, the Robust Posterior Sampling variant PS Robust$_N$ also utilizes $N$ samples.

\begin{table}[h]
	\centering
	\begin{tabular}{|p{4cm}|l|l|}
		\hline
		Guidance & Normalized & Unormalized \\
		\hline
		$mean(R^2)$  &0.902 & 0.966 \\ 
			\hline
	\end{tabular}
	\caption{Reported $R^2$ values are derived from FDTD simulations conducted after inverse design using DSB PS Robust$_5$.}
	\label{table::PS_robust}
\end{table}

	Robust posterior sampling, akin to Monte Carlo posterior sampling, introduces substantial computational demands during the optimization of schedules $g_t$ and $s(t)$, as well as throughout the generation phase. Both generation time and memory usage scale linearly with the number of samples, mirroring the computational overhead observed in MC posterior sampling. 
	\subsection{Amplitude-constrained guidance}
Existing posterior sampling methods, including Monte Carlo and Robust Posterior Sampling, focus on accurately estimating the guidance term or incorporating robustness within the framework of Jensen's approximation. Their primary goal is to compute the most precise and relevant guidance term. However, once obtained, this guidance term essentially a gradient is applied without explicit constraints or criteria.

The choice of the guidance coefficient, which directly affects the overall amplitude of the guidance term, is critical. Performance varies significantly between normalized and unnormalized guidance terms, as evidenced by prior posterior sampling techniques such as raw posterior sampling, Monte Carlo posterior sampling, and Robust Posterior Sampling. The conclusions in \cite{chung2022diffusion} regarding the optimal guidance coefficient are empirical and likely problem-dependent. For instance, in the metasurface inverse design context, unnormalized guidance consistently yields better results.

Even when the guidance term is accurately computed to mitigate Jensen's approximation limitations \cite{song2023loss}, it may still violate statistical constraints, as shown in \cite{yang2024guidance}. The subsequent posterior sampling method leverages the inherent amplitude constraints on the guidance term imposed by the DSB framework, as demonstrated in \cite{yang2024guidance}.
	\subsubsection{Spherical gaussian constraint posterior sampling}
	 A constrained posterior sampling method, based on the Spherical Gaussian (SG) constraint, was introduced in \cite{yang2024guidance}. Unlike the two previously discussed posterior sampling techniques, this approach dynamically adjusts the amplitude of the guidance term at each step of the noise sampling process, while retaining the direction derived from standard posterior sampling. This approach introduces a lower bound on Jensen's gap, motivating the need to constrain the guidance term. Since Jensen's gap scales with the dimensionality, the guidance term error increases with the dimension of $x$. To address this, the direction of the guidance term is preserved while its amplitude is adjusted to satisfy the Spherical Gaussian constraint.
	
		While the reverse process updates for DMs and DSBs differ (see Equation~\eqref{reverse_update} in Appendix~\ref{App:maths} and the corresponding formulation in \cite{grand2026enhanced}), the analogous hypersphere constraining $x_{t-1}$ in DSBs is defined as
	\[
	\mathbb{S}_{\mu_\theta(x_t, c, t), \sqrt{n} \, \Sigma_{t,t-1}}^{n},
	\]
	where $n$ represents the dimensionality of $x_t$ and $\Sigma_{t,t-1}$ is specified in Equation~\eqref{reverse_update} (Appendix~\ref{App:maths}). The amplitude is determined by the Spherical Gaussian constraint, whereas stochastic perturbations may be applied to the direction to promote solution diversity. Additional details are available in \cite{yang2024guidance}.
	
	The reverse process update incorporating the Spherical Gaussian constraint is then given by:
	
	\begin{equation}
		\label{predicted_mean_SG}
		\mu_\theta^{ps}(x_t, c, t) = \mu_\theta(x_t, c, t) - \sqrt{n}\Sigma_{t,t-1}\frac{\nabla_{x_t} \| c  - S_\phi(x_{0|t}) \|^2_2}{\|\nabla_{x_t} \| c - S_\phi(x_{0|t}) \|^2_2\|_2},
	\end{equation}
	
From a practical perspective, SG eliminates the need for empirical tuning of the guidance coefficient $q_t$. Applying SG to raw posterior sampling significantly improves performance, as demonstrated in Table~\ref{table}.

\begin{figure}[H]
		\centering
		\includegraphics[scale=0.55]{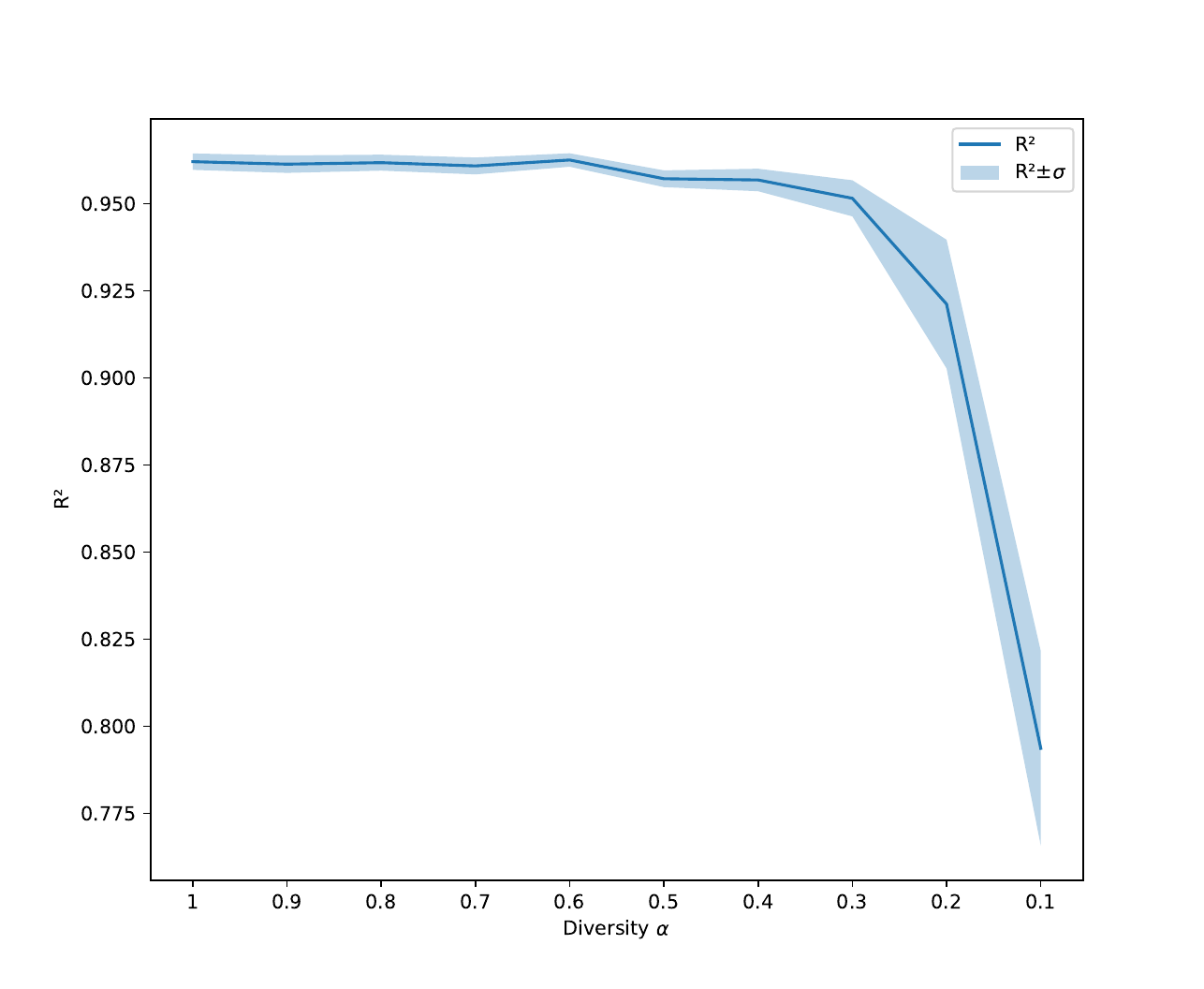}
		\caption{The $R^2$ metric, computed after simulation of inverse-designed metasurface parameters using Spherical Gaussian Constrained Posterior Sampling, is evaluated for varying magnitudes of the diversity coefficient $\alpha$. Remarkably, the $R^2$ remains high even at low values of $\alpha$ and stays stable until the gradient-based guidance contributes less than half to the overall update.}
		\label{fig:diversity_r2}
	\end{figure}

When adding direction stochasticity for diversity Equation \eqref{predicted_mean_SG} becomes:
\begin{equation}
		\mu_\theta^{ps}(x_t, c, t) = \mu_\theta(x_t, c, t) - \sqrt{n}\Sigma_{t,t-1}\frac{G}{\|G\|_2}, \label{eq:dps_sg_diversity}
\end{equation}
with $G = \alpha\frac{\nabla_{x_t} \| c  - S_\phi(x_{0|t}) \|^2_2}{\|\nabla_{x_t} \| c - S_\phi(x_{0|t}) \|^2_2\|_2}+ (1-\alpha)\frac{\epsilon}{\|\epsilon\|_2},  \alpha \in [0,1] $
Following the formulation in \cite{yang2024guidance}, the relative influence of diversity controlled by the coefficient $g_r$ is ambiguous, as it weights two vectors with differing norms. To address this, Equation \eqref{eq:dps_sg_diversity} normalizes both terms prior to weighting by $\alpha$, ensuring a balanced contribution. As it is shown in Figure \ref{fig:diversity_r2} the performance for DSB SG remains high and consistent until a low $\alpha$ so a high value of stochasticity. The performance only starts to decline when the gradient guidance term represents less than 30\% of the total guidance term (gradient and stochasticity). This methodology facilitates the generation of highly diverse solutions without compromising peak performance. A visual examination of outputs produced with different $\alpha$ values does not reveal any discernible pattern in diversity. Additional details are provided in Appendix~\ref{App:diversity}.

	\subsubsection{Disk gaussian constraint posterior sampling}
	We propose two extensions to the SG constraint from \cite{yang2024guidance}, which are equally applicable to DMs. While SG constrains the guidance term amplitude within the high-confidence region of the unconditional intermediate sample $x_t$, specifically on the hypersphere 
	$\mathbb{S}_{\mu_\theta(x_t, c, t), \sqrt{n} \, \Sigma_{t,t-1}}^{n},$
	it assumes that the solution lies on this surface. If $x_t$ resides inside the Hypersphere rather than on its boundary, SG tends to overestimate the guidance amplitude. Such overshoot degrades the precision of posterior sampling both during intermediate steps and in the final output.
	
	To address this, it is essential to explore guidance amplitudes within the volume enclosed by the hypersphere, especially in the later sampling stages where $x_t$ approaches a solution inside the confidence region. Excessively large guidance amplitudes at this stage can be detrimental.
	
	We introduce the Disk Gaussian (DG) constraint, which samples the guidance amplitude such that the guidance term lies within the hyperdisk 
	$\mathbb{D}_{\mu_\theta(x_t, c, t), \sqrt{n} \, \Sigma_{t,t-1}}^{n}.$
	Specifically, the guidance amplitude $r_{DG}$ is drawn from a uniform distribution,
	$r_{DG} \sim \mathcal{U}(0, \Sigma_{t,t-1}),$
	allowing exploration of the interior volume of the confidence region. As reported in Table~\ref{table}, this hyperdisk exploration achieves performance comparable to PS SG on a $23 \times 23$ pillar metasurface.
	
	Early in the generation process, when the current estimate is distant from the optimum, uniformly sampling guidance amplitudes in $\mathcal{U}(0, \Sigma_{t,t-1})$ can be counterproductive, as small guidance steps may hinder progress. This interpretation is corroborated by scaling results in Section~\ref{sec:scaling}, where PS DG underperforms relative to methods that constrain early steps near the boundary of the confidence interval, specifically the hypersphere 
	$\mathbb{S}_{\mu_\theta(\mathbf{x}_t, c, t), \sqrt{n} \, \Sigma_{t,t-1}}^{n}.$

	\subsubsection{Ring Gaussian constraint posterior sampling}
We introduce a hybrid posterior sampling constraint, termed the Ring Gaussian (RG) constraint, which integrates features of both the Spherical Gaussian and Disk Gaussian constraints. In the DG approach, the guidance amplitude is uniformly sampled from $\mathcal{U}(0, \Sigma_{t,t-1})$, allowing for small amplitudes during early generation steps when the estimate is distant from the solution. However, this method underexploits the larger amplitudes allowed by the hypersphere radius $\Sigma_{t,t-1}$.

	To address this, we introduce a generation-step-dependent radius interval that uniformly samples the guidance amplitude within a hyper ring. This hyper ring is defined by a center and two radii: an outer radius $r_{\mathrm{out}}$ fixed to the hypersphere radius $\Sigma_{t,t-1}$, and an inner radius 
	$r_{\mathrm{in}} = \Sigma_{t,t-1} (1 - \mu_{0|t}),$
	which starts near the hypersphere boundary in early steps and decreases to zero, converging to a disk in later steps.
	
	As demonstrated in Figure~\ref{table}, the RG constraint achieves the highest average performance alongside SG constraint and Robust Posterior sampling. RG’s advantage becomes particularly pronounced when scaling to larger metasurfaces, where its constraints outperform all other methods as shown in Section \ref{sec:scaling}.
	
Building upon the limitations of stochastic guidance reduction in PS DG and PS RG relative to PS SG, a promising research direction involves integrating the dynamic gradient amplitudes from prior steps into the posterior sampling framework. By adapting principles akin to those in the ADAM optimizer \cite{adam2014method}, this approach seeks to refine guidance steps based on observed convergence behavior, potentially enhancing the robustness and efficiency of posterior sampling guidance in inverse design.

	\section{Comparative results} \label{sec:results}
	\subsection{Diffusion steps} \label{sec:steps_study}
\begin{figure}[H]

		\hspace{-4cm}
		\includegraphics[scale=0.48]{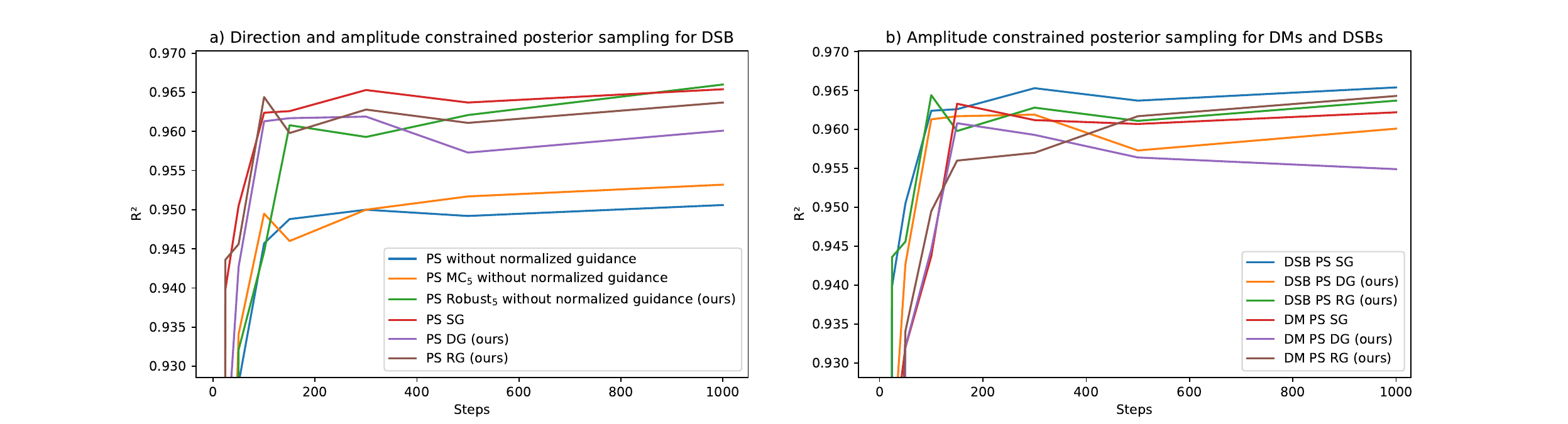}
		\caption{a) $R^2$ values for DSBs using direction and amplitude-constrained posterior sampling. b) Comparative performance analysis of amplitude-constrained posterior sampling between DSBs and DMs.}
		\label{fig:steps_study_duo}
	\end{figure}

	An important SBs and DMs parameter is the number of diffusion steps, the more steps there are, the more computationnaly and time expensive it will be. The stakes of generation with these diffusion steps is to reduce them as much as possible while maintaining high performance. Hence, on Figure \ref{fig:steps_study_duo} the behavior of different posterior sampling techniques addressed in this paper are analyzed. As shown on Figure \ref{fig:steps_study_duo}a, among the different posterior techniques addressed, SG, DG and RG are the one performing the best by reaching the higher $R^2$ for the least steps. To reduce even more then number of generating steps, techniques such as distillation \cite{luo2023latent} and consistency models \cite{song2023consistency} have been developed to build few steps generators with DMs and similar techniques can be used with DSBs. However, these methods have not yet been explored for DSBs.  
	
	Figure~\ref{fig:steps_study_duo}b demonstrates that DSBs and DMs achieve similar performance under amplitude-constrained posterior sampling. Notably, the figure reveals distinct behavioral patterns in posterior sampling between DMs and DSBs as the number of generation steps varies, with DSBs converging to plateau $R^2$ values more efficiently, requiring fewer steps than DMs.

	\subsection{Inverse Design Precision}
 
This section evaluates the performance of DMs and DSBs under a fixed sampling budget of 1000 steps, with the Gradient Descent reference method allocated an equivalent number of iterations. As shown in Table~\ref{table}, DSBs employing PS Robust$_5$ achieve the highest average performance across all targets, followed closely by DSBs with amplitude constraints SG and RG. Visual representations of the far field are provided in Appendix~\ref{App:comp}.

	\begin{table}[h]
		\centering
		\begin{tabular}{|p{5cm}|l|l|}
			\hline
			Method & $mean(R^2)$ & $std(R^2)$ \\
				\hline
			DSB PS without normalized guidance  &0.951 & 0.009  \\ 
				\hline
			DSB PS MC$_5$ without normalized guidance &0.953 & 0.010  \\ 
				\hline
			\textbf{DSB Robust$_5$ without normalized guidance (ours)}  &\textbf{0.966} & \textbf{0.007 } \\ 
				\hline
			DSB SG  &0.965 & 0.007  \\ 
				\hline
				DSB DG (ours) & 0.960& 0.015  \\ 
				\hline
			DSB RG (ours) & 0.964 & 0.009  \\ 
			\hline
			
			DM SG & 0.962&0.006  \\ 
			\hline
			Phase Retrieval \& Local Model &0.913&0.020   \\ 
			\hline
			Gradient Descent initialized on $\mathcal{U}(r_{min},r_{max})$ & 0.815&0.009   \\ 
			\hline
		\end{tabular}
		\caption{On average, DSB with Robust Posterior Sampling significantly outperforms Gradient Descent initialized on $\mathcal{U}(r_{\min}, r_{\max})$ as well as the Phase Retrieval and Local Model methods, while achieving a modest improvement over diffusion models employing PS with SG and RG constraints.}
		\label{table}
	\end{table}
	\subsection{Integrating Enhanced Guidance Estimation with Amplitude Constraints}
The guidance term can be decomposed into two distinct components: direction and amplitude. Guidance directions may be derived using raw posterior sampling, Monte Carlo posterior sampling, or Robust posterior sampling. Following the determination of the guidance direction, various amplitude constraints such as SG, DG, or RG can be applied.

As evidenced by the comparative analysis in Table~\ref{table} and Table~\ref{table:cross}, the application of amplitude constraints to enhanced posterior sampling methods results in only marginal performance variations. While the precision gains are modest, a consistent pattern is observed: integrating amplitude constraints (SG or RG) with guidance computed via Monte Carlo or Robust posterior sampling yields superior performance relative to their unconstrained versions. Nevertheless, these outcomes remain outperformed by the direct application of amplitude constraints to raw posterior sampling, underscoring the pivotal role of amplitude constraints.

This suggests that errors arising from Jensen's approximation which Monte Carlo posterior sampling aims to mitigate exert a comparatively minor effect relative to the impact of amplitude constraints, as initially proposed in \cite{yang2024guidance}.
		\begin{table}[h]
		\centering
		\begin{tabular}{|p{5cm}|l|l|}
			\hline
			Method & $mean(R^2)$ & $std(R^2)$ \\
			\hline
			DSB PS MC$_5$ \& SG  &0.959 & 0.008  \\ 
			\hline
			DSB PS MC$_5$ \& RG (ours)  &0.962 & 0.009  \\ 
			\hline
			\textbf{DSB Robust$_5$ (ours) \& SG}  &\textbf{0.963} & \textbf{0.006}  \\ 
			\hline
			DSB Robust$_5$ (ours) \& RG (ours) &0.960 & 0.007  \\ 
			\hline
		
		\end{tabular}
		\caption{On average, there is no significant advantage of combining different guidance term computation schemes with different guidance amplitude constraints.}
		\label{table:cross}
	\end{table}

	\subsection{Robustness}

	Robustness of solutions is critical to mitigate the effects of small manufacturing variations in metasurface fabrication, as highlighted in the Robust Posterior Sampling section. To quantify robustness, we analyze similarly to \cite{yao2018hessian}, the Hessian matrix of the loss function with respect to the metasurface parameters. Utilizing automatic differentiation frameworks such as PyTorch, the Hessian is efficiently computed. Formally, the Hessian is the matrix of second-order partial derivatives:
	$\frac{\partial^2 \mathrm{Loss}}{\partial x_i \partial x_j}, \quad (i,j) \in \llbracket1, \ldots, N\rrbracket,$
	where $N$ is the number of metasurface parameters.
	
\begin{figure}[H]
		\centering
		\includegraphics[scale=0.60]{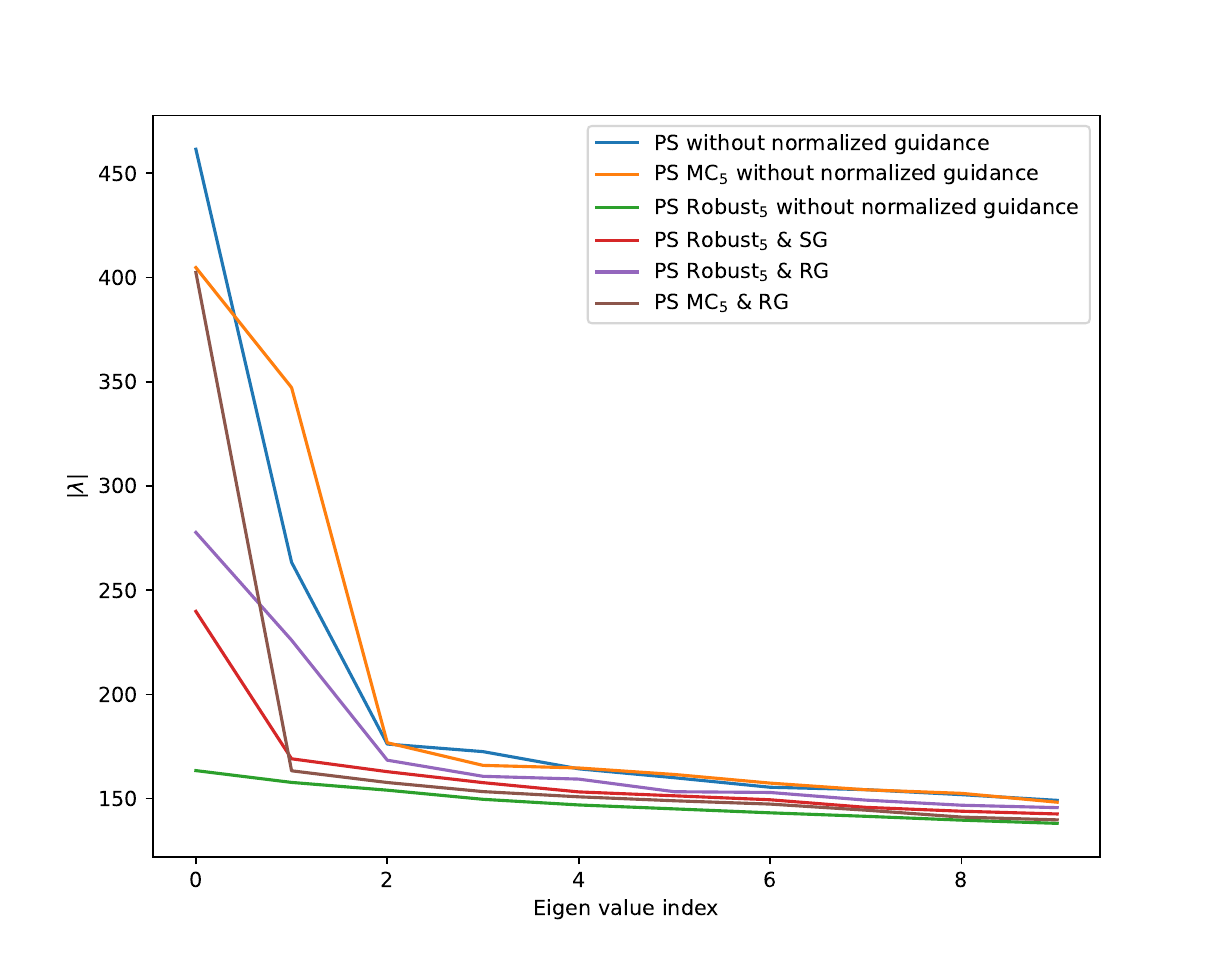}
		\caption{Spectrum of the Hessian for the three highest-performing and three lowest-performing posterior sampling schemes under direction constraints, amplitude constraints, and their combined application.}
		\label{fig:robustness}
	\end{figure}

	The Hessian characterizes the local curvature of the loss landscape and thus the sensitivity of the solution to input perturbations. A robust solution corresponds to a flat region where all second derivatives approach zero:
	$\forall (i,j)  \in \llbracket1, \ldots, N\rrbracket, \quad \frac{\partial^2 \mathrm{Loss}}{\partial x_i \partial x_j} \approx 0.$
To quantify flatness, we solve the Hessian eigenvalue problem and order the eigenvalues in descending magnitude. Figure~\ref{fig:robustness} presents the first ten eigenvalues, where the largest eigenvalue corresponds to the direction of maximum curvature. For comparison across posterior sampling methods, we focus on this principal curvature direction. Consistent with theoretical expectations, Robust Posterior Sampling produces the smallest largest eigenvalue, indicating significantly more robust inverse designs.

	\subsection{Scaling} \label{sec:scaling}
\begin{figure}[H]

\hspace{-4cm}
			\includegraphics[scale=0.48]{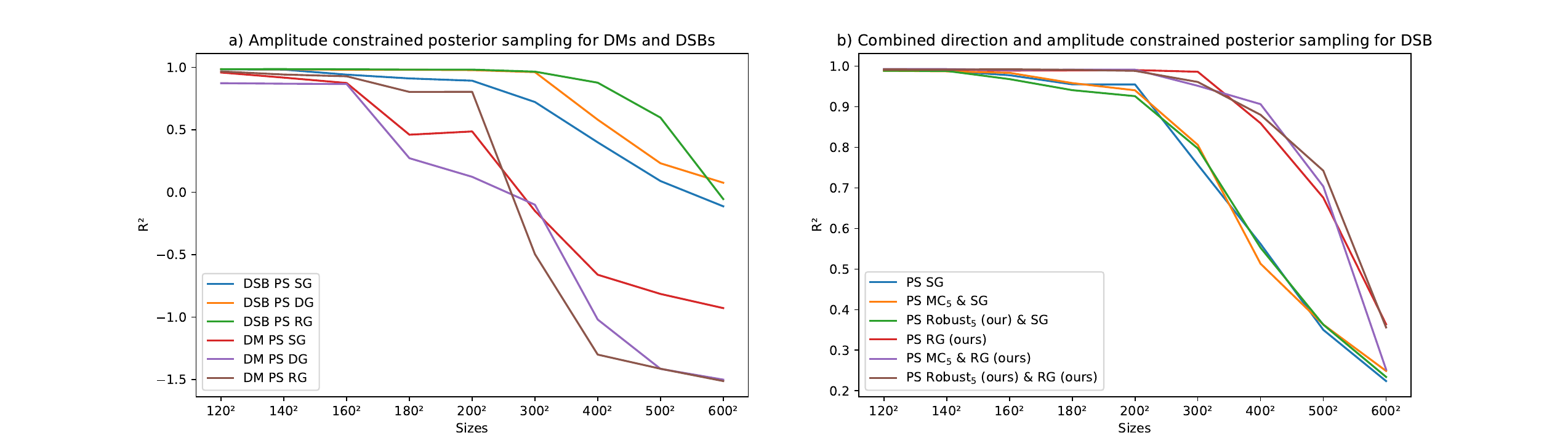}
			\caption{a) Comparison of $R^2$ values for the far field magnitude, as inferred by the surrogate model $S_\phi$, across metasurfaces inverse-designed using DMs and DSBs with Direction and Amplitude-Constrained Posterior Sampling. While DM performance degrades significantly for metasurfaces exceeding $120 \times 120$ pillars, DSBs sustain high performance up to $350 \times 350$ pillars, demonstrating scalability to metasurfaces nearly an order of magnitude larger.
				b) Illustration of the minimal impact of DSBs direction posterior sampling when combined with amplitude-constrained posterior sampling.}
			\label{fig:giga_scaling_duo}
		\end{figure}

	As highlighted in \cite{grand2026enhanced}, a key limitation of prior work is the exclusive focus on small metasurfaces consisting of $23 \times 23$ pillars. This size was selected due to computational constraints associated with building the simulation database. Specifically, FDTD simulations are computationally intensive, and constructing a database of 5,000 samples within a two-week timeframe necessitated this compromise. Nevertheless, the ultimate objective is to extend inverse design methodologies to larger metasurfaces. This section investigates the scalability of DSBs to increasingly larger metasurfaces. Performance is first evaluated on metasurface sizes amenable to tractable simulations and no performance difference is shown between DMs and DSBs. For metasurfaces exceeding $100 \times 100$ pillars, Figure~\ref{fig:giga_scaling_duo}a reveals a pronounced performance disparity between DSBs and DMs. Across all evaluated posterior sampling methods, DSBs systematically surpass DMs, attaining superior $R^2$ values and sustaining performance over increased metasurface dimensions prior to degradation. The far field amplitudes derived from inverse design at these scales are detailed in Appendix~\ref{App:scaling}. In cases where direct simulation is computationally infeasible, the surrogate model $S_\phi$ is utilized to approximate the $R^2$ performance metric.
	
		Among the posterior sampling techniques applied to DSBs, DG and RG methods introduced here surpass SG method from \cite{yang2024guidance}. Both DG and RG sustain a high $R^2$ plateau up to metasurfaces of size $300 \times 300$ pillars, whereas performance with SG begins to decline at smaller scales. Consequently, DSBs combined with RG demonstrate the best overall performance among all methods evaluated.
		
	When integrating various guidance computation schemes (Raw, MC, Robust) with distinct constraints (RG, SG), the outcomes exhibit a strong dependence on the choice of constraint, while the influence of the guidance direction computation scheme remains comparatively minor, as illustrated in Figure~\ref{fig:giga_scaling_duo}b. Visual representations of the far field are provided in Appendix~\ref{App:scaling}.

			\FloatBarrier
	\section{Conclusion}
	
	This study explores posterior sampling strategies tailored for Diffusion Schr\"odinger Bridges in the context of metasurface inverse design. Initially applied to small $23 \times 23$ pillar metasurfaces using Monte Carlo posterior sampling ($R^2 = 0.953$), we introduce a Robust posterior sampling method that improves both performance ($R^2 = 0.966$) and solution robustness. Building upon foundational techniques such as the Spherical Gaussian constraint ($R^2 = 0.956$), we propose novel variants, Disk Gaussian ($R^2 = 0.960$) and Ring Gaussian constraints ($R^2 = 0.964$) which refine guidance term calibration and enable more effective exploration within confidence regions.
	
	Our analysis demonstrates that these advanced sampling methods enhance stability and performance, particularly for large-scale metasurfaces exceeding the capabilities of conventional diffusion models. Notably, the Disk and Ring Gaussian constraints are the only methods to sustain superior $R^2$ metrics for metasurfaces scaling up to $300 \times 300$ pillars, underscoring their robustness in high-dimensional inverse design challenges, whereas the performance of Diffusion Models declines significantly for metasurfaces larger than $120 \times 120$ pillars.
	
	Furthermore, the integration of deterministic and stochastic guidance amplitude schedules provides a versatile framework to balance precision and diversity in the generated solutions.
	
	Overall, this work establishes posterior sampling as a critical component in leveraging DSBs for complex inverse design tasks, providing both theoretical insights and practical algorithms that advance the state of the art. Future research may explore adaptive guidance schedules and further surrogate model integration to extend applicability to even larger-scale and more diverse physical systems.
	
	\clearpage
	\bibliography{references.bib}
	\bibliographystyle{unsrturl}
	\onecolumn
	\appendix
\section{Reference Methods} \label{App:ref}

\subsection{Phase Retrieval \& Local model}

A core challenge in metasurface design arises from the limited availability of far field amplitude data, which precludes the direct retrieval of near-field information through inverse Fourier transformation. To address this limitation, phase retrieval algorithms \cite{jaganathan2016phase,wang2017hybrid} are utilized to approximate the near-field phase profile. This estimated phase is subsequently translated into meta-atom parameters via a bijective mapping. However, the combined phase retrieval and local model approach relies on two critical assumptions: (1) the transmission properties of meta-atoms are independent of their parameters, and (2) interactions between meta-atoms are negligible. Consequently, this method is constrained to metasurfaces characterized by predominantly local effects \cite{isnard2024advancing} and is inapplicable to complex structures, such as stacked metasurfaces, where inter-layer coupling plays a significant role.

By contrast, inverse design methodologies based on DMs or SBs circumvent these restrictive assumptions. They demonstrate efficient scalability to metasurfaces with non-local effects and a high number of parameters per meta-atom, thereby providing a more versatile and robust solution.

\subsection{Gradient Descent}

Gradient descent stands as a foundational optimization algorithm in machine learning, designed to minimize a loss function by iteratively refining model parameters. This loss function quantifies the deviation between predicted and target outputs. The parameter update is governed by the rule:

\begin{equation}
	\theta \leftarrow \theta - \alpha \nabla_{\theta} L(\theta),
\end{equation}

where $\alpha$ represents the learning rate, dictating the magnitude of each update, and $\nabla_{\theta} L(\theta)$ is the gradient of the loss function $L(\theta)$ with respect to the parameters $\theta$.

In the context of metasurface design, the parameters $\theta$ define the metasurface configuration, while the loss function evaluates the discrepancy between the desired far field amplitude and the simulated far field response derived from the current configuration. To streamline optimization, a neural network-based surrogate solver replaces computationally demanding methods like FDTD, enabling rapid simulation and gradient computation due to its fully differentiable nature.

Leveraging modern deep learning frameworks such as Jax and PyTorch, automatic differentiation efficiently computes gradients by decomposing complex functions into elementary operations through a computational graph. Backpropagation traverses this graph in reverse, calculating the gradients of the loss with respect to each parameter. These gradients inform parameter updates via optimization algorithms, including Stochastic Gradient Descent (SGD) \cite{amari1993backpropagation} or  ADAM \cite{adam2014method} .

Initially developed to train deep neural networks, automatic differentiation has been adapted to differentiable solvers \cite{kim2023torcwa,ponomareva2025torchgdm} and surrogate solvers \cite{lim2022maxwellnet,gao2019bidirectional}. These advancements have established gradient-based optimization as a powerful tool in metasurface design and broader scientific applications.

As demonstrated in \cite{grand2026enhanced}, metasurface inverse design for far field distribution with GD exhibits heightened sensitivity to the $0^{th}$ diffraction order and energy-concentrating diffraction orders, resulting in reduced precision. To mitigate this issue, two supplementary terms were incorporated into the loss function, specifically targeting the $0^{th}$ order and the diffraction orders collecting energy. The inclusion of these terms improved precision, increasing the $R^2$ value from $0.697$ to $0.815$.
\section{Mathematical formulation}\label{App:maths}
\subsection{Diffusion Models}

Diffusion models are a class of generative models that synthesize data by progressively denoising a sample starting from pure Gaussian noise. The generative process involves two main stages: a forward diffusion process, which gradually adds Gaussian noise to the data over a sequence of time steps, and a reverse denoising process, which learns to reconstruct the original data by iteratively removing noise. 

Mathematically, the forward process defines a Markov chain that transforms data $x_0$ into a noisy version $x_T$ through a sequence of conditional Gaussian distributions. The reverse process approximates the posterior distribution $p_\theta(x_{t-1}|x_t)$ parameterized by a neural network, enabling the generation of high-fidelity samples by sequentially denoising from $x_T$ back to $x_0$. 

Training optimizes the network parameters by minimizing a variational bound on the negative log-likelihood, often implemented as a simplified denoising score matching objective. Diffusion models have demonstrated superior sample quality and training stability compared to other generative frameworks, making them well-suited for complex inverse design tasks.
	\subsection{Schr\"odinger Bridge Mathematical Formulation}
	In this section, we briefly present the mathematical formulation of DMs to establish a parallel with Diffusion Bridges (SBs). This comparison facilitates the discussion of methods and results from \cite{grand2026enhanced} on DMs in the context of DSBs.
	
	Given data $X_0$ sampled from a distribution $p_A$, DMs can be defined as the following Stochastic Differential Equation (SDE) \cite{ song2020score} of the form : 
	\begin{equation}
		dX_t = f_t(X_t) dt + \sqrt{\beta_t} dW_t,
		\label{equ:DM_fSDE}
	\end{equation}
	where the diffusion coefficient $\beta_t \in \mathbb{R}$ is appropriately chosen and the drift $f_t$ is linear in $X_t$. The terminal distribution at $t=1$ converges to a normal distribution, $X_1 \sim \mathcal{N}(0, I)$. 
	
	Reversing this forward SDE yields the time-reversed SDE \cite{anderson1982reverse}:
	\begin{equation}
		dX_t = \bigl[f_t(X_t) - \beta_t \nabla \log p(X_t, t)\bigr] dt + \sqrt{\beta_t} d\bar{W}_t,
		\label{equ:DM_bSDE}
	\end{equation}
	where $p(\cdot, t)$ denotes the marginal density of the forward process at time $t$ and $\nabla \log p$ is its score function. This reversed SDE shares the same marginal distributions as the forward SDE and its path measure coincides almost surely with that of the forward process.
	
	The connection to diffusion generative models is established through the following forward and backward stochastic differential equations defining SBs:
	\begin{subequations}
		\begin{align}
			dX_t &= \bigl[f(X_t) +  \beta_t \nabla \log \Psi(X_t,t)\bigr] dt +  \sqrt{\beta_t}  dW_t, \label{equ:SB_fSDE}\\
			dX_t &= \bigl[f(X_t) - \beta_t \nabla \log \hat{\Psi}(X_t,t)\bigr] dt +  \sqrt{\beta_t}  d\bar{W}_t,
		\end{align}
	\end{subequations}
	where the wave functions $\Psi$ and $\hat{\Psi}$ satisfy the partial differential equations:
	\begin{align}
		\frac{\partial \Psi(X_t,t)}{\partial t} &= - \nabla \Psi(X_t,t)^\top f(X_t,t) - \frac{1}{2} \beta_t \Delta \Psi(X_t,t), \label{eq:diff1} \\
		\frac{\partial \hat{\Psi}(X_t,t)}{\partial t} &= - \nabla \cdot \bigl(\hat{\Psi}(X_t,t) f(X_t,t)\bigr) + \frac{1}{2}\beta_t \Delta \hat{\Psi}(X_t,t),
	\end{align}
	with boundary conditions $\Psi(x,0) \hat{\Psi}(x,0) = p_A(x)$ and $\Psi(x,1) \hat{\Psi}(x,1) = p_B(x)$ for densities $p_A$ and $p_B$.
	
	Compared to Equation~\eqref{equ:DM_fSDE}, Equation.~\eqref{equ:SB_fSDE} includes an additional nonlinear drift term, $\beta_t \nabla \log \hat{\Psi}(x,t)$, which enables diffusion between distributions that are not necessarily standard normal at the process endpoint. Furthermore, $\nabla \log \hat{\Psi}(x,t)$ is related to, but distinct from, the score function of the perturbed data, since
	\begin{equation}
		\Psi(x_t,t) \hat{\Psi}(x_t,t) = q(x_t,t),
	\end{equation}
	implying
	\begin{equation}
		\nabla \log \Psi(x_t,t) + \nabla \log \hat{\Psi}(x_t,t) = \nabla \log q(x_t,t).
	\end{equation}
	
	Although these equations resemble those of standard diffusion generative models, deriving a general solution remains challenging. The authors of \cite{liu20232} introduce a tractable approach, I2SB, based on paired observations satisfying $p(x_a, x_b) = p_A(x_a) p_B(x_b|x_a)$. This assumption is valid for metasurface inverse design, where pairs of metasurface structures and their corresponding electromagnetic fields are obtained via FDTD simulations \cite{gedney2011introduction}.
	
	Under this approximation and by setting the linear drift $f(x,t) := 0$, the posterior $q(x|x_a, x_b)$ admits the analytic Gaussian form \cite{liu20232}:
	\begin{equation}
		q(x|x_a, x_b) = \mathcal{N}\bigl(x_t; \mu_t(x_a, x_b), \Sigma_t \bigr),
	\end{equation}
	with
	\begin{equation}
		\mu_t = \frac{\bar{\sigma}_t^2}{\bar{\sigma}_t^2 + \sigma_t^2} x_a + \frac{\sigma_t^2}{\bar{\sigma}_t^2 + \sigma_t^2} x_b, \quad
		\Sigma_t = \frac{\sigma_t^2 \bar{\sigma}_t^2}{\bar{\sigma}_t^2 + \sigma_t^2} \cdot I,
	\end{equation}
	
	\begin{equation}
		\mu_t = \mu_a(t) x_a + \mu_b(t) x_b, 
	\end{equation}
	where $\sigma_t^2 := \int_0^t \beta(\tau) d\tau$ and $\bar{\sigma}_t^2 := \int_t^1 \beta(\tau) d\tau$. Given pairs $(x_a, x_b)$, one can sample $x_t$ as
	\begin{equation}
		x_t = \mu_t + \Sigma_t \epsilon, \quad \epsilon \sim \mathcal{N}(0, I),
	\end{equation}
	for any $t \in [0,1]$.
	
	Applying Tweedie's formula \cite{efron2011tweedie} to Equation~\eqref{eq:diff1} yields
	\begin{equation}
		x_0 = x_t - \frac{\log \hat{\Psi}(x_t, t)}{\sigma_t}.
	\end{equation}
	Accordingly, the loss function is defined as
	\begin{equation}
		\mathcal{L}_{\mathrm{I2SB}} = \frac{1}{2} \mathbb{E}_{x, t, \epsilon} \left\| \epsilon_\theta(x_t, t) - \frac{x_t - x_0}{\sigma_t} \right\|_2^2.
	\end{equation}
	where $\|.\|_2$ the right term approximates the score function of the backward drift $\nabla \log \hat{\Psi}(x,t)$, which is then used during sampling to transport samples from $p_B$ to $p_A$.
	
	Sampling is conducted via standard recursive methods like DDPM \cite{ho2020denoising}, where the denoised estimate at step $t < T$ is given by
	
	\begin{equation}
		p_\theta(x_{t-1} | x_t) = \mathcal{N}(x_{t-1}; \mu_\theta(x_t, t), \sigma_t^2 {I}),
	\end{equation}
	
	where the predicted mean $\mu_\theta(x_t, t)$ is computed as:
	\begin{equation}
		\mu_\theta(x_t, t) = \mu_{0|t}x_{0|t} + \mu_tx_t \label{eq:predicted_mean}
	\end{equation}
	with : 
	\begin{subequations}
		\begin{align}
			x_{0|t} = x_t- \epsilon_\theta(x_t, t) \sigma_t \label{posterior_mean}\\
			\mu_{0|t} =\frac{\sigma_{t-1}^2}{\sigma_t^2}  \\
			\mu_t =\frac{\sigma_t^2 -\sigma_{t-1}^2}{\sigma_t^2}
		\end{align}
	\end{subequations}
	The reverse process update rule is given by
	\begin{equation}
		x_{t-1} = \mu_\theta(x_t, t) + \Sigma_{t,t-1} \epsilon, \label{reverse_update}
	\end{equation}
	where 
	$\Sigma_{t,t-1} = \frac{\sigma_{t-1} \sqrt{\sigma_t^2 - \sigma_{t-1}^2}}{\sigma_t}.$
	Enhanced recursive sampling methods \cite{chung2022diffusion, song2023loss, yang2024guidance} can be employed to further improve performance.
	
	\section{Training improvements} \label{App:training}
	This section presents two methods designed to enhance the performance of DSBs. As noted in \cite{grand2026enhanced}, although these methods impact training metrics, such metrics alone do not suffice to identify the optimal training strategy. Instead, the selection criterion relies on the final performance metric computed from the far field response obtained via inverse design followed by rigorous FDTD simulation. Throughout this paper, the term \emph{final metric} refers to the evaluation performed on the far field response after inverse design and simulation.
	
	The conclusions drawn herein utilize the improved sampling techniques described in the next section. Notably, the ancestral sampling method introduced in \cite{ho2020denoising} and applied to DSBs proves ineffective for the inverse design problem considered.

	The two training improvements, previously addressed for DMs in \cite{grand2026enhanced}, are analyzed here for their impact on DSBs. 
	\subsection{Noising schedule} \label{App:schedule}
	As demonstrated in \cite{nichol2021improved,grand2026enhanced}, the choice of noise schedule significantly influences DM performance depending on the application. To optimize far field precision, we introduced three distinct schedules for the noise variance $\beta(t)$, a monotonically increasing function satisfying $\beta(0) = \beta_{\text{start}}$ and $\beta(T) = \beta_{\text{end}}$. Detailed results and definitions of these schedules are provided bellow. It highlights two key observations consistent with \cite{grand2026enhanced}. First, the choice of noising schedule significantly affects the final performance metric, making its optimization essential for achieving optimal results. Second, a favorable training metric does not necessarily translate to superior final performance. Hence, DSBs show the same kind of sensitivity to noising schedule than DMs.
	
	Three distinct scheduling functions for the noise variance $\beta_t$ are introduced. These functions define the noise addition schedule used during the diffusion process.
	\begin{itemize}
		\item Linear schedule : $\beta_t = \beta_{\text{start}} + t(\beta_{\text{end}}-\beta_{\text{start}})$
		\item Quadratic schedule : $\beta_t = (\sqrt{\beta_{\text{start}}} + t(\sqrt{\beta_{\text{end}}}-\sqrt{\beta_{\text{start}}}))^2$
		\item Sigmoid schedule : $\beta_t = \beta_{\text{start}} + \frac{1}{1+e^{t}}(\beta_{\text{end}}-\beta_{\text{start}})$
	\end{itemize}
	
\begin{figure}[h!]
		\centering
		\includegraphics[scale=0.43]{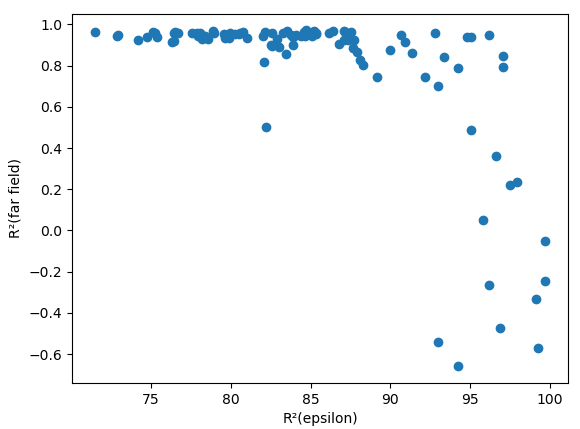}
		\caption{Each blue dot represents a diffusion Schr\"odinger bridge trained with distinct schedule parameters sampled from the set
			$\{\text{schedule function}\} \times \{\beta_{\mathrm{end}}\} \times \{\text{consistency loss}\} = \{\text{sigmoid}, \text{quadratic}, \text{linear}\} \times \{0.1, 0.01, 0.001\} \times \{\text{no consistency}, \text{scheduled consistency}, \text{consistency}\}.$
			When combined with enhanced sampling techniques, the best training metric does not necessarily correspond to the best final performance.
		}
		\label{schedule_results}
	\end{figure}

The final performance exhibits significant variability depending on the scheduling function and its parameters, $\beta_{\text{start}}$ and $\beta_{\text{end}}$, as illustrated in Figure~\ref{schedule_results}.
		
		\subsection{Consistency Loss}\label{App:cons_loss}
		
		The consistency loss, introduced in \cite{grand2026enhanced} for DMs, reduces their sensitivity to diffusion hyperparameters, thereby alleviating the optimization burden.
		
		This loss term enforces the constraint $A(x) = c$. In our context, $A$ corresponds to solving Maxwell's equations via FDTD; however, for computational efficiency, the FDTD solver is replaced by a surrogate model $S_\phi$. During training, $x_{0|t}$ computed from Equation~\eqref{posterior_mean} is used to evaluate the consistency loss, denoted as $\mathcal{L}_{\mathrm{consistency}}$.
		The overall loss function is defined as
		\begin{equation}
			\mathcal{L} = \mathcal{L}_{\mathrm{diff}}\big(\epsilon_\theta(x_t, c, t), \epsilon \big) + \gamma(t) \, \mathcal{L}_{\mathrm{consistency}}\big(S_\phi(x_{0|t}), c\big).
			\label{weighted_consistency_loss}
		\end{equation}
		
		Let $x_{0|t}$ be defined by Equation~\eqref{posterior_mean}, $\mathcal{L}_{\mathrm{diff}}$ denote the diffusion loss, and $\mathcal{L}_{\mathrm{consistency}}$ quantify the discrepancy between the surrogate output $S_\phi(x_{0|t})$ and the conditioning variable $c$. As in DMs, incorporating the consistency loss reduces the sensitivity of DSBs to hyperparameters, thereby decreasing performance variability and improving overall outcomes. This approach was used to efficiently train the SDB model evaluated in the following experiments.
		
\begin{figure}[H]
		\centering
		\includegraphics[scale=0.45]{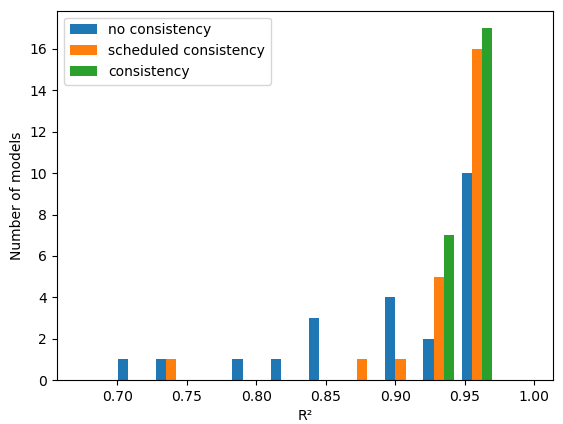}
		\caption{Final performance metrics are presented for the three consistency strategies employed during training: no consistency loss, standard consistency loss, and scheduled consistency loss. The results demonstrate that the application of consistency loss to DSBs yields a tighter concentration of outcomes near the optimal value.}
		\label{fig:consistency_loss}
	\end{figure}

		 The weighting factor $\gamma(t)$ can be constant, e.g., $\gamma(t) = 1$, as illustrated in Figure \ref{fig:consistency_loss} for the fixed consistency loss. Alternatively, $\gamma(t)$ may vary with $t$, such as $\gamma(t) = \mu_a(t)$, which approaches 1 as $t \to 0$, corresponding to the scheduled consistency loss shown in Figure \ref{fig:consistency_loss}.

	\section{Posterior sampling} \label{App:posterior_sampling}
\begin{figure}[H]
		\centering
		\includegraphics[scale=0.5]{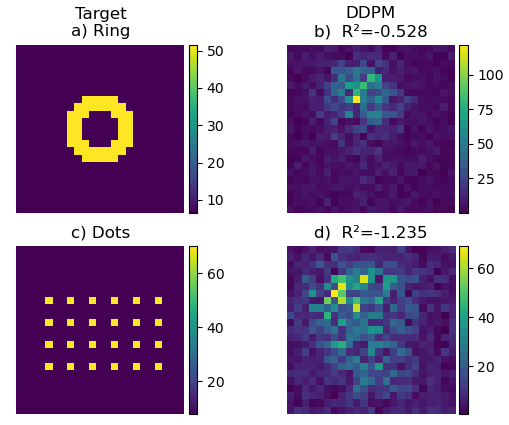}
		\caption{The far field patterns obtained from simulations of inverse-designed metasurface parameters using ancestral sampling with 1000 steps are shown in panels (b) and (d), compared to the corresponding target patterns in panels (a) and (c).}
		\label{fig:ddpm_results}
	\end{figure}

	Solely conditioning the model resulted in unsatisfactory performance. Representative far field examples are presented in Figure \ref{fig:ddpm_results}. To overcome this limitation, a guidance term is introduced during sampling, resulting in a posterior sampling method (PS). Analogous to the consistency loss, this approach utilizes Tweedie's formula \cite{efron2011tweedie} to compute the guidance term from the posterior mean $x_{0|t}$. Unlike the standard score $\nabla_{x_t} \log \hat{\Psi}(x_t)$, posterior sampling explicitly incorporates the condition by evaluating $\nabla_{x_t} \log p(x_t \mid c)$, which by Bayes' rule decomposes as
	\begin{equation}
		\nabla_{x_t} \log \hat{\Psi}(x_t \mid c) = \nabla_{x_t} \log \hat{\Psi}(c \mid x_t) + \nabla_{x_t} \log \hat{\Psi}(x_t).
	\end{equation}
	
	The term $\nabla_{x_t} \log \hat{\Psi}(c \mid x_t)$ is intractable and is approximated using Jensen's inequality. For a random variable $x \sim \hat{\Psi}(x)$ and function $f$, the approximation states:
	\begin{equation}
		\mathbb{E}_{x \sim \hat{\Psi}(x)}[f(x)] \approx f\big(\mathbb{E}_{x \sim \hat{\Psi}(x)}[x]\big).
	\end{equation}
	Applying this to posterior sampling yields
	\begin{equation}
		\hat{\Psi}(c \mid x_t) \approx \hat{\Psi}(c \mid x_{0|t}),
	\end{equation}
	where $\hat{\Psi}(c \mid x_{0|t})$ is evaluated using an electromagnetic simulator approximated by the neural network $S_\phi$, i.e., $S_\phi(x_{0|t}) = c$. Consequently, the guidance term is approximated as
	\begin{equation}
		\nabla_{x_t} \log \hat{\Psi}(c \mid x_t) \approx - \nabla_{x_t} \| c - S_\phi(x_{0|t}) \|^2_2.
	\end{equation}
	
	The following observations are applicable to both conditional and unconditional variants of DSBs. Consequently, the predicted mean $\mu_\theta$ in the update rule of Equation~\eqref{reverse_update} is adjusted as follows: 
	\begin{equation}
		\label{predicted_mean_posterior_sampling}
		\mu_\theta^{ps}(x_t, c, t) = \mu_\theta(x_t, c, t) 
		- \nabla_{x_t} \| c - S_\phi(x_{0|t}) \|^2_2,
	\end{equation}
	
	\subsection{Ancestral sampling} \label{App:ancestral}
	
	\subsection{Raw posterior sampling} \label{App:raw_PS}
	
	\begin{figure}[H]
		\centering
		\includegraphics[scale=0.5]{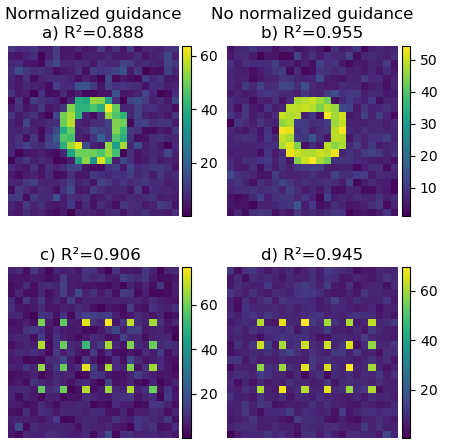}
		\caption{$|$Far field$|$ obtained via simulation of inverse-designed metasurface parameters using DSB PS over 1k steps. Normalized guidance is defined as $q_t = \frac{1}{\|c - S_\phi(x_{0|t})\|_2}$, while non-normalized guidance is set to $q_t = 1$}
		\label{fig:dps_results}
	\end{figure}
	\subsection{Monte Carlo posterior sampling}\label{App:PS_MC}
	\cite{song2023loss} demonstrates, using a one-dimensional Gaussian example, that the guidance term computed via Tweedie's formula combined with Jensen's approximation lacks sufficient accuracy. The guidance term is being overestimated for large $\sigma_t$ and underestimated for small $\sigma_t$. To mitigate this, they propose sampling around the posterior mean and averaging to reduce the estimation error. Accordingly, the predicted mean $\mu_\theta$ is modified as
	\begin{equation}
		\mu_\theta^{mc}(x_t, c, t) = \mu_\theta(x_t, c, t) \\
		+ \nabla_{x_t} \log \left( \frac{1}{N} \sum_{i=1}^N \exp \big(-\| c - S_\phi(x_{0|t}^i) \|^2_2 \big) \right),
	\end{equation}
	Let $x_{0|t}^i$ be sampled from 
	$\mathcal{N}\big(x_{0|t}, g_t^2 \mathbf{I}\big).$
	
\begin{figure}[H]
		\centering
		\includegraphics[scale=0.5]{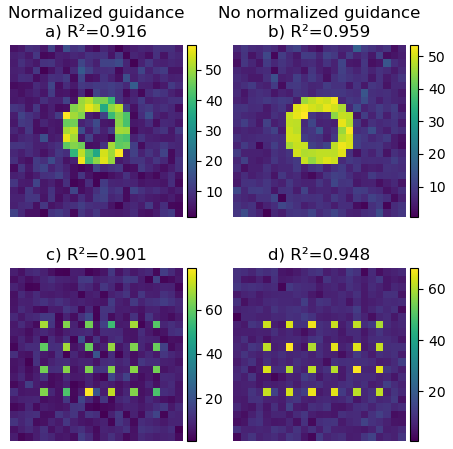}
		\caption{$|$Far field$|$ obtained via simulation of inverse-designed metasurface parameters using DSB PS MC$_5$ over 1k steps. Normalized guidance is defined as $q_t = \frac{1}{\|c - S_\phi(x_{0|t})\|_2}$, while non-normalized guidance is set to $q_t = 1$}
		\label{fig:dps_mc_results}
	\end{figure}

	This posterior sampling method incorporates an additional schedule, $g_t$, which regulates the extent of Monte Carlo sampling. The optimization process for $g_t$ is detailed bellow, which also includes illustrative far field examples on Figure \ref{fig:dps_mc_results}. Although Monte Carlo posterior sampling with $N$ samples (PS MC$_N$) enhances guidance term estimation in the Gaussian scenario, as illustrated in \cite{song2023loss}, its performance in the metasurface inverse design problem addressed in this study remains analogous to that of raw posterior sampling, with no clear improvements. Similar to raw posterior sampling, Monte Carlo posterior sampling degrades performance when using normalized guidance terms as shown in Table \ref{table::PS_MC}. Therefore, only results for Monte Carlo posterior sampling without normalized guidance are presented in Table \ref{table}. This behavior of diffusion Schr\"odinger bridges closely parallels that observed for diffusion models in \cite{grand2026enhanced}.
	
	Importantly, Monte Carlo posterior sampling incurs substantial computational overhead, both in optimizing $g_t$ and during generation, as generation time and memory consumption scale linearly with the number of samples $N$.
	
	\begin{table}[H]
		\centering
		\begin{tabular}{|p{4cm}|l|l|}
			\hline
			Guidance & Normalized & Unormalized \\
			\hline
			$mean(R^2)$  &0.897 & 0.953  \\ 
			\hline
		\end{tabular}
		\caption{Reported $R^2$ values are derived from FDTD simulations conducted after inverse design using DSB PS MC$_5$.}
		\label{table::PS_MC}
	\end{table}
	
\begin{figure}[H]
		\centering
		\includegraphics[scale=0.5]{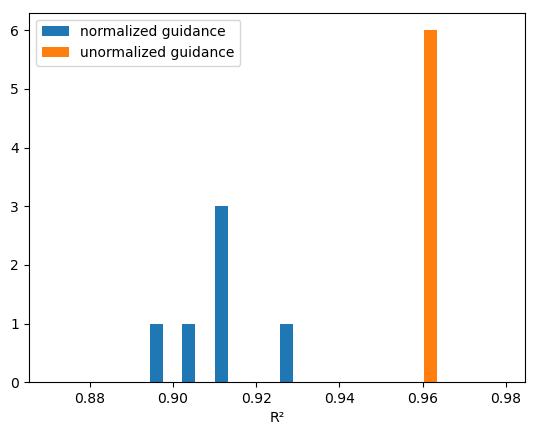}
		\caption{Final metric results for various Monte Carlo posterior sampling schedules are reported under both normalized and unnormalized guidance. Unnormalized guidance consistently outperforms normalized guidance and exhibits robustness to the choice of the $g(t)$ schedule.}
		\label{fig:dps_mc_hist}
	\end{figure}

Several strictly decreasing Monte Carlo posterior sampling schedules $g_t$ have been evaluated to enable broad sampling in early steps, gradually focusing samples near the mean in later steps. The schedules considered include:
\begin{itemize}
	\item Linear schedule: \quad $g_t = \frac{1 - \mu_{0|t}}{C}$
	\item Quadratic schedule: \quad $g_t = \frac{(1 - \mu_{0|t})^2}{C}$
\end{itemize}
where $C \in \{1, 50, 100\}$. The performance of these schedules is illustrated in Figure~\ref{fig:dps_mc_hist}.
	
	\subsection{Robust posterior sampling}\label{App:PS_Robust}
	
		\begin{figure}[H]
		\centering
		\includegraphics[scale=0.5]{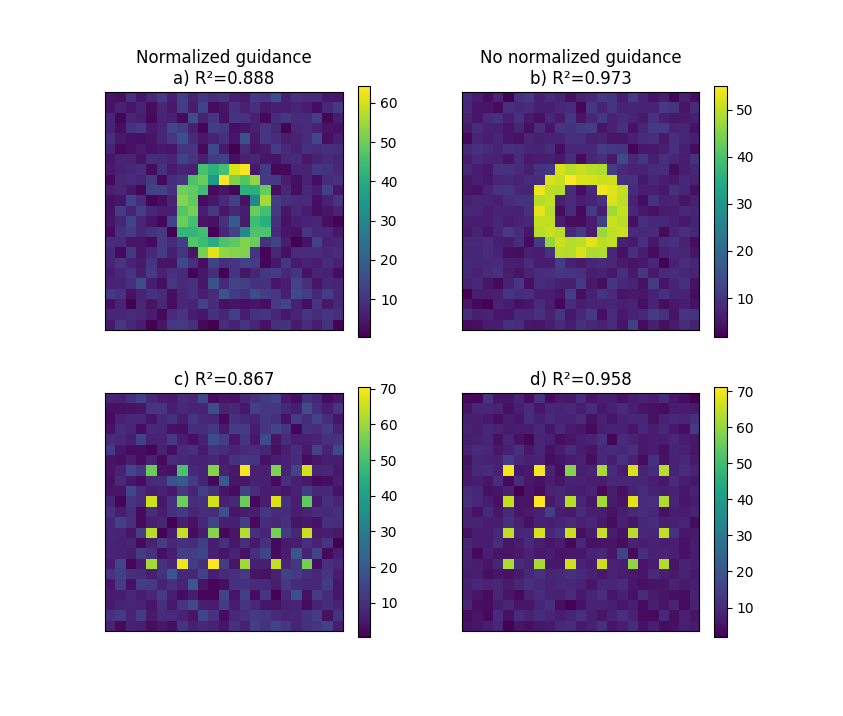}
		\caption{$|$Far field$|$ obtained via simulation of inverse-designed metasurface parameters using DSB PS Robust$_5$ over 1k steps. Normalized guidance is defined as $q_t = \frac{1}{\|c - S_\phi(x_{0|t})\|_2}$, while non-normalized guidance is set to $q_t = 1$}
		\label{fig:dps_robust_results}
	\end{figure}
	
	\begin{figure}[H]
		\centering
		\includegraphics[scale=0.5]{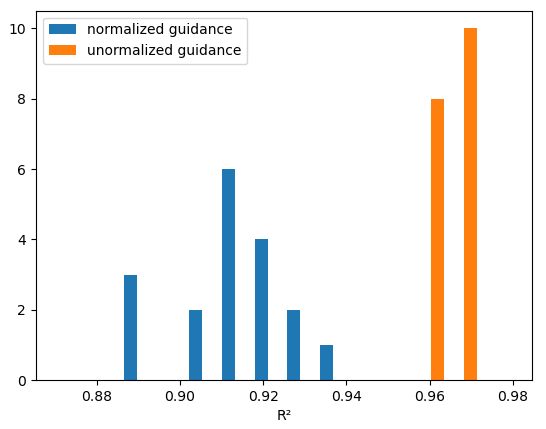}
		\caption{Final metric results for various Robust posterior sampling schedules are reported under both normalized and unnormalized guidance. Unnormalized guidance consistently outperforms normalized guidance and exhibits robustness to the choice of the $g(t)$  and $s(t)$ schedule.}
		\label{fig:dps_robust_hist}
	\end{figure}

	\subsection{Spherical gaussian constrained posterior sampling}\label{App:PS_SG}

		\begin{figure}[H]
		\centering
		\includegraphics[scale=0.5]{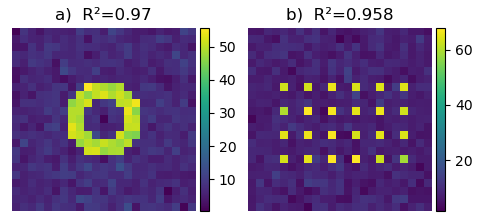}
		\caption{$|$Far field$|$ obtained via simulation of inverse-designed metasurface parameters using DSB SG over 1k steps with $\alpha=1$.}
		\label{fig:dps_sg_results}
	\end{figure}
	
	\begin{figure}[H]
		\centering
		\includegraphics[scale=0.6]{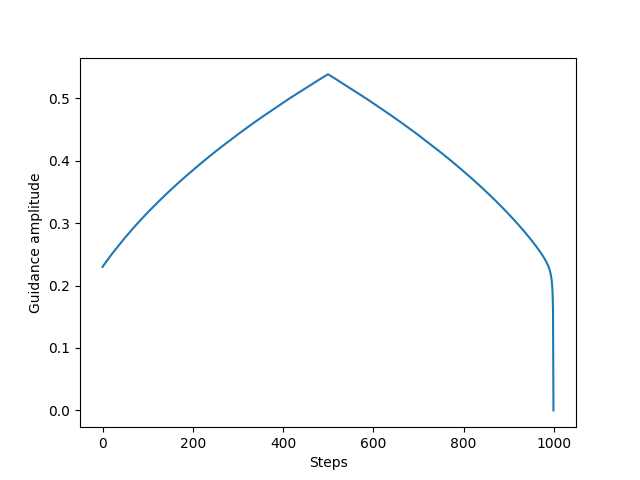}
		\caption{Guidance term amplitude for SG}
		\label{fig:r_SG}
	\end{figure}
	\subsection{Disk gaussian constrained  posterior sampling} \label{App:PS_DG}
		\begin{figure}[H]
		\centering
		\includegraphics[scale=0.5]{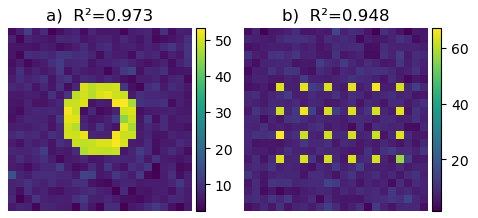}
		\caption{$|$Far field$|$ obtained via simulation of inverse-designed metasurface parameters using DSB DG over 1k steps with $\alpha=1$.}
		\label{fig:dps_dg_results}
	\end{figure}

		\begin{figure}[H]
		\centering
		\includegraphics[scale=0.6]{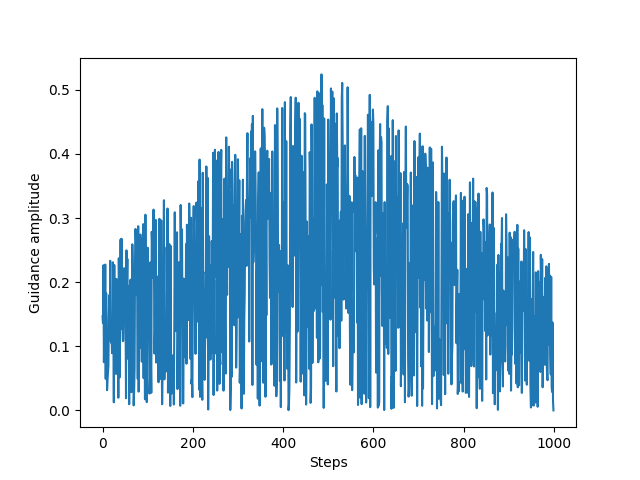}
		\caption{Guidance term amplitude for DG}
		\label{fig:r_DG}
	\end{figure}

	\subsection{Ring gaussian constrained  posterior sampling} \label{App:PS_RG}
	
		\begin{figure}[h]
		\centering
		\includegraphics[scale=0.5]{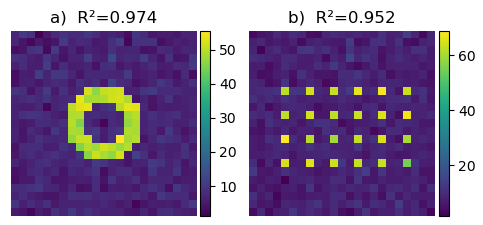}
		\caption{$|$Far field$|$ obtained via simulation of inverse-designed metasurface parameters using DSB RG over 1k steps with $\alpha=1$.}
		\label{fig:dps_rg_results}
	\end{figure}

	\begin{figure}[H]
		\centering
		\includegraphics[scale=0.6]{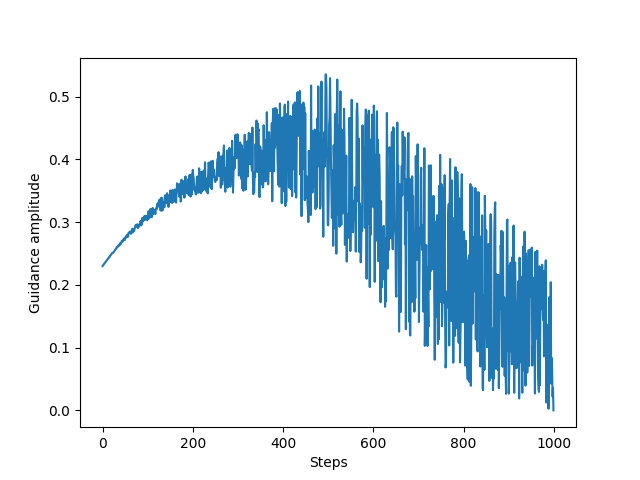}
		\caption{Guidance term amplitude for RG}
		\label{fig:r_RG}
	\end{figure}
	\subsection{Reduced Spherical gaussian constrained posterior sampling} \label{App:PS_rSG}
	
\begin{figure}[H]
		\centering
		\includegraphics[scale=0.48]{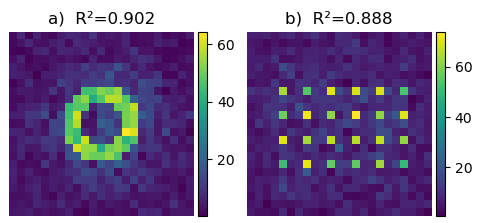}
		\caption{$|$Far field$|$ obtained via simulation of inverse-designed metasurface parameters using DSB reduced SG over 1k steps with $\alpha=1$.}
		\label{fig:dps_reduced_sg_results}
	\end{figure}

	Rather than introducing stochasticity in the guidance amplitude via uniform sampling as in DG and RG methods, we propose a deterministic guidance amplitude schedule, referred to as the reduced Spherical Gaussian (rSG) method. As emphasized in \cite{yang2024guidance}, the upper bound imposed by the SG constraint is essential to prevent divergence from the intermediate manifold $\mathcal{M}_t$. However, maintaining the guidance amplitude constantly at this upper bound may impede convergence when the distance between the true solution $\mathbf{x}_0$ and the predicted mean $\mu_\theta(\mathbf{x}_t, t)$ is smaller than the amplitude. Since this distance is intractable due to the unknown exact solution, the Reduced Spherical Gaussian constraint proposes restricting the guidance amplitude to a fraction of the hypersphere. Specifically, the solution is constrained to lie on the hypersphere
	$\mathbb{S}_{\mu_\theta(\mathbf{x}_t, c, t), \sqrt{n} \,(1 - \mu_{0|t}) \Sigma_{t,t-1}}^{n},$
	thereby allowing an additional gradual reduction of the guidance amplitude toward the end of the generation process. The radius schedule closely corresponds to the RG method, with the reduced hypersphere radius $r_{rSG}(t)$ matching the inner radius defined in RG. As illustrated in Figure~\ref{fig:dps_reduced_sg_results}, the rSG method underperforms relative to other posterior sampling techniques. This outcome suggests that merely adhering to the SG constraint and reducing guidance amplitude without accounting for the dynamic evolution of guidance across steps is inadequate. Consequently, a promising avenue for future research involves dynamically adjusting the guidance amplitude based on previous steps, while respecting the upper bound imposed by the SG constraint.
	
	\begin{figure}[H]
		\centering
		\includegraphics[scale=0.6]{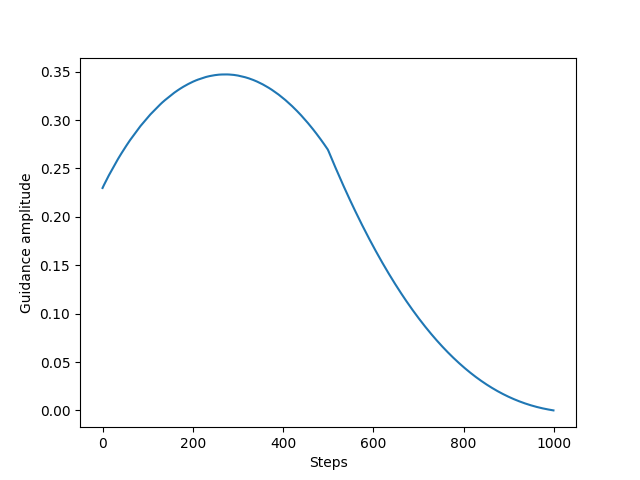}
		\caption{Guidance term amplitude for reduced SG}
		\label{fig:r_rSG}
	\end{figure}
	\section{Diversity} \label{App:diversity}
\begin{figure}[H]
		\centering
		\includegraphics[scale=0.45]{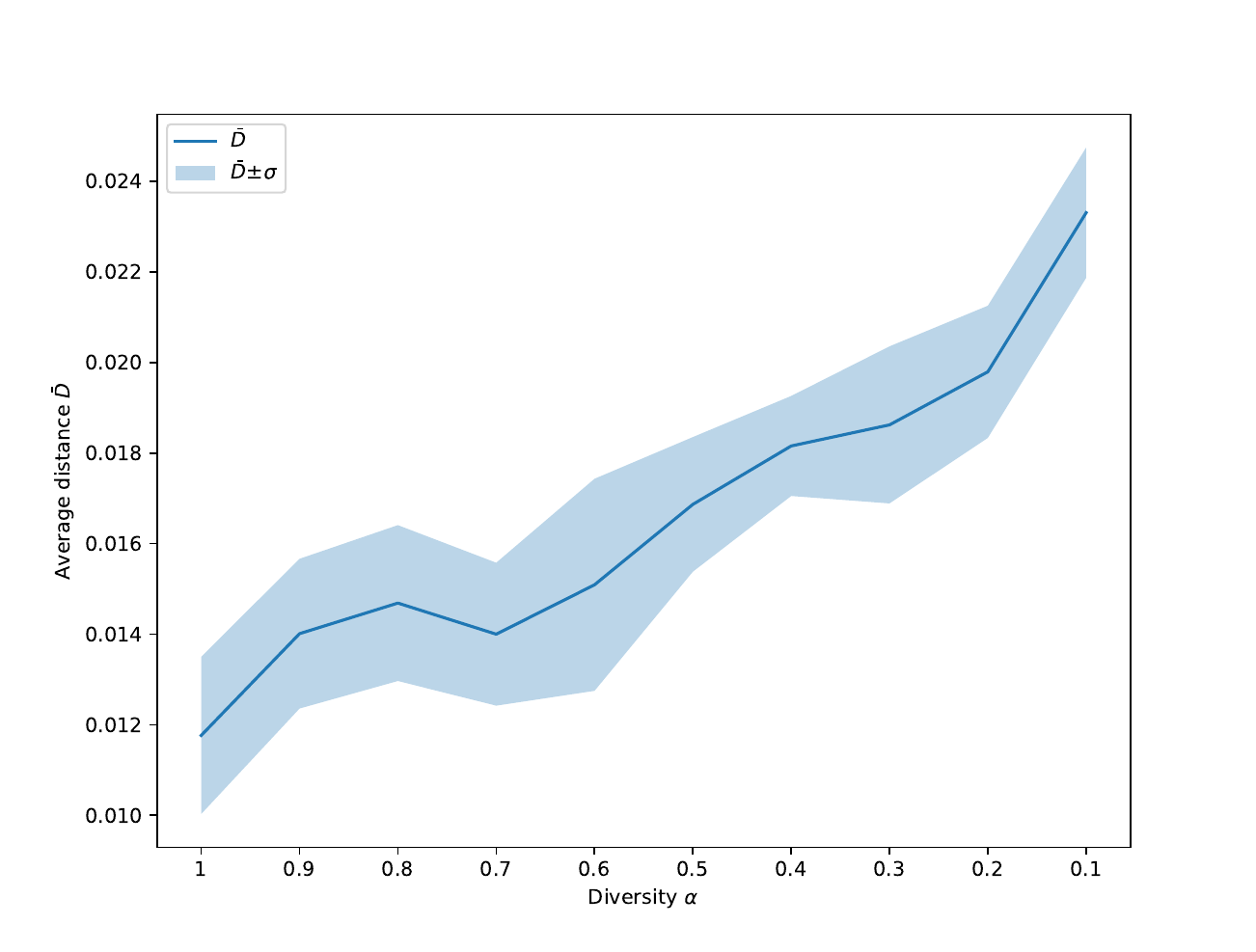}
		\caption{The diversity metric $D$ and its standard deviation, computed over ten samples generated with varying diversity coefficients $\alpha$ using Spherical Gaussian Constrained Posterior Sampling, are evaluated after simulating inverse-designed metasurface parameters.}
		\label{fig:diversity_distance}
	\end{figure}

\begin{figure}[H]
	\hspace{-1cm}
	\includegraphics[scale=0.3]{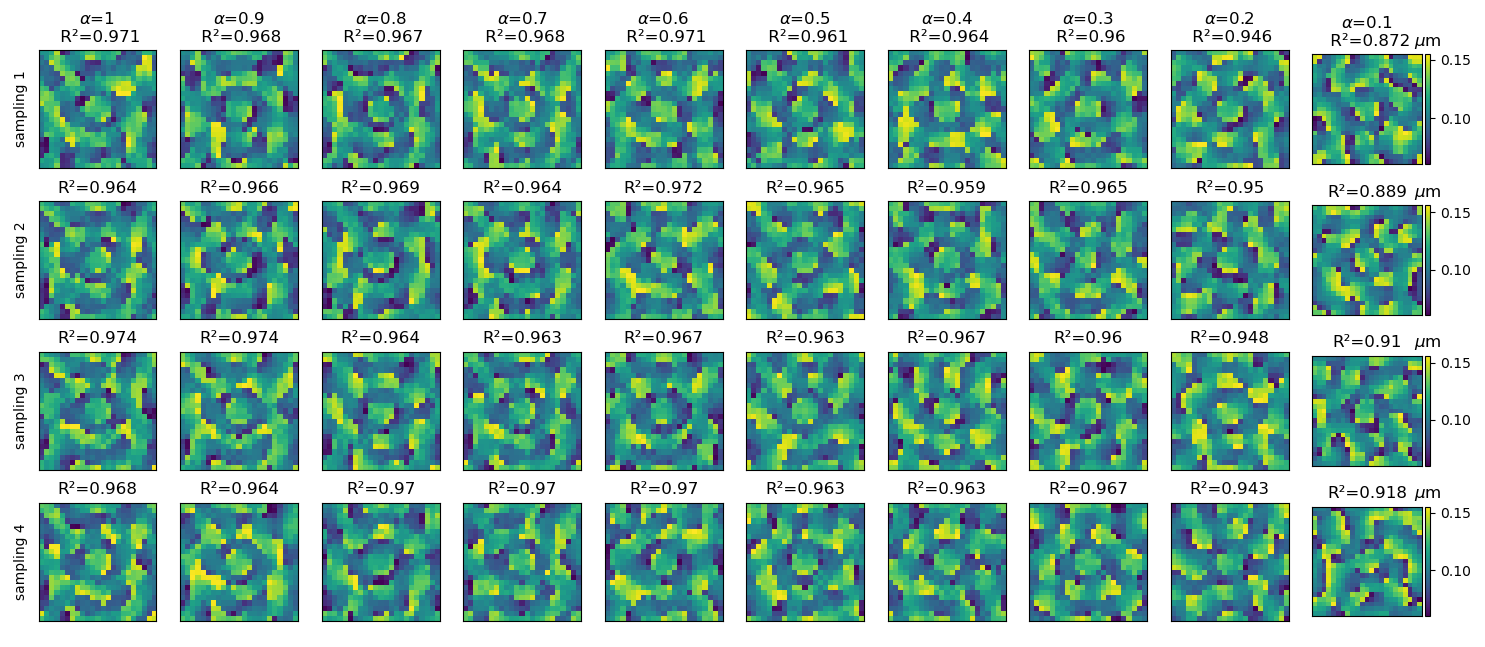}
	\caption{Metasurface parameters generated for smaller and smaller diversity coefficient $\alpha$.}
	\label{fig:r_diversity}
\end{figure}

	To quantify the diversity of metasurface parameters, we introduce the metric $\bar{D}$.
	\begin{multline}
		\bar{D}_\alpha = \frac{1}{\#\Omega}\sum_{(i,j) \in \Omega}\|P_i-P_j\|_1, \\ \Omega = \{(i,j) \in \mathbb{N}^2: i \neq j, i \le N_\alpha \text{ and } j \le N_\alpha\}
	\end{multline}
	With $P_i$ representing $i^{th}$ metasurface with $d$ parameters after inverse design, $N_\alpha$ the number of inverse design done for a given diversity coefficient $\alpha$ and $\|.\|_1$ is a Mean Absolute deviation.
	As illustrated in Figure~\ref{fig:diversity_distance}, decreasing $\alpha$ leads to increased solution diversity, evidenced by a rise in the average pairwise distance between samples. Specifically, reducing $\alpha$ from 1 to 0.3 nearly doubles this average distance $\sum_{(i,j) \in \Omega}$  while preserving comparable $R^2$ performance.
	Figure \ref{fig:r_diversity} displays the initial four samples employed in the computation of the diversity metric $\bar{D}$.
	
\section{Comparative results far fields}\label{App:comp}
		\begin{figure}[H]
	\hspace{-3.5cm}
	\includegraphics[scale=0.4]{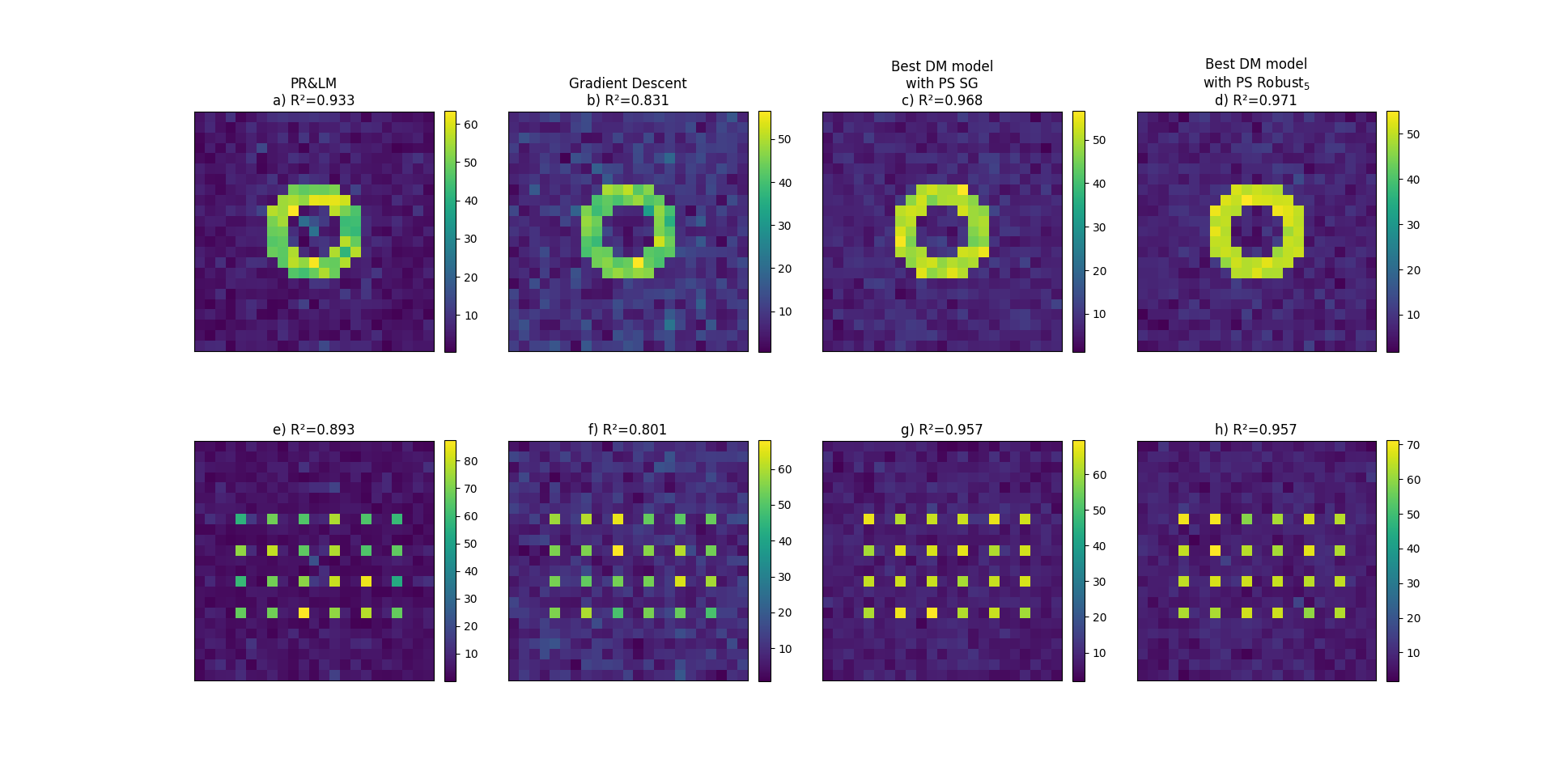}
	\caption{$|$Far field$|$ after simulation from inverse-designed metasurface parameters using, Phase Retrieval \& Local model (a, e), Gradient Descent initialized on $\mathcal{U}(r_{min},r_{max})$ (b, f),  DM PS SG (c, g),  DM PS Robust$_5$ (d, h).}
	\label{fig:method_comparison_ref}
\end{figure}

\begin{figure}[H]
	\centering
	\includegraphics[scale=0.43]{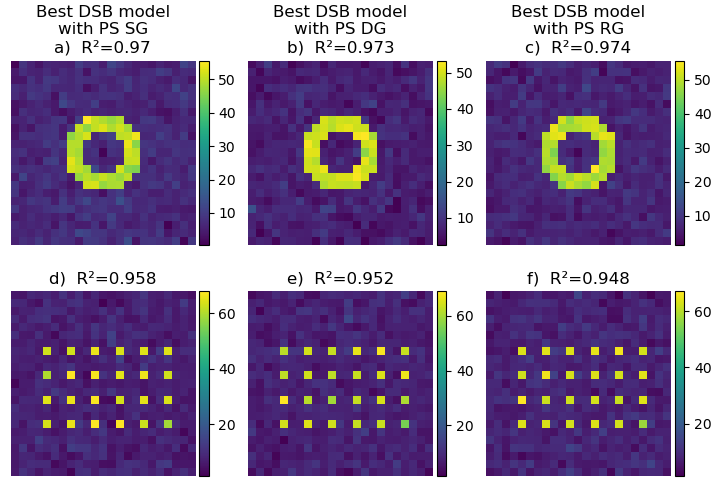}
	\caption{$|$Far field$|$ after simulation from inverse-designed metasurface parameters using, best DSB  with PS SG (a, d), best DSB with PS DG (b, e), best DSB with PS RG (c, f).}
	\label{fig:method_comparison_best}
\end{figure}

	\section{Scaling} \label{App:scaling}

\begin{figure}[h!]
	
		\hspace{-4cm}
		\includegraphics[scale=0.5]{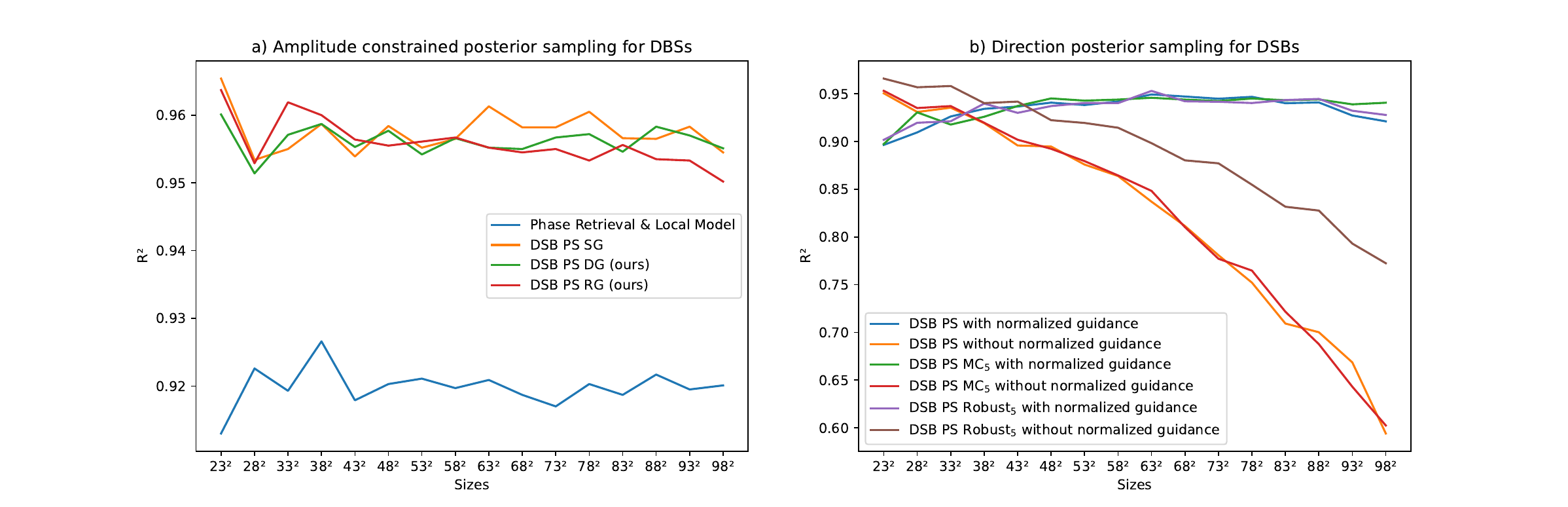}
		\caption{a) Comparison of $R^2$ values for simulated far field magnitudes obtained from inverse-designed metasurfaces of increasing size using Diffusion Schr\"odinger Bridges (DSBs) with amplitude-constrained posterior sampling techniques (SG, DG, RG) against the Phase Retrieval \& Local Model method.
			b) Performance of alternative posterior sampling direction techniques (Raw, MC, Robust), including the option for normalized guidance.}
		\label{fig:scaling_duo}
	\end{figure}

	Empirical results indicate that guidance term normalization adversely affects performance in PS, PS MC, and PS Robust methods, as evidenced by Figures~\ref{fig:dps_results}, \ref{fig:dps_mc_results}, and~\ref{fig:dps_robust_results}. However, when applied to large-scale metasurface inverse design, normalization yields a marginal improvement in the $R^2$ metric. In contrast, PS, PS MC, and PS Robust methods without normalized guidance exhibit a pronounced performance decline, as illustrated in Figure~\ref{fig:scaling_duo}b.
	
	Additionally, Figure~\ref{fig:scaling_duo}b presents a comparative analysis of DSBs utilizing amplitude-constrained posterior sampling methods SG, DG, and RG against the Phase Retrieval \& Local Model reference method. All amplitude-constrained posterior sampling approaches sustain consistently high performance across the largest metasurface dimensions amenable to simulation, surpassing the Phase Retrieval \& Local Model technique.
	
	\begin{figure}[H]
			\hspace{-4cm}
		\includegraphics[scale=0.45]{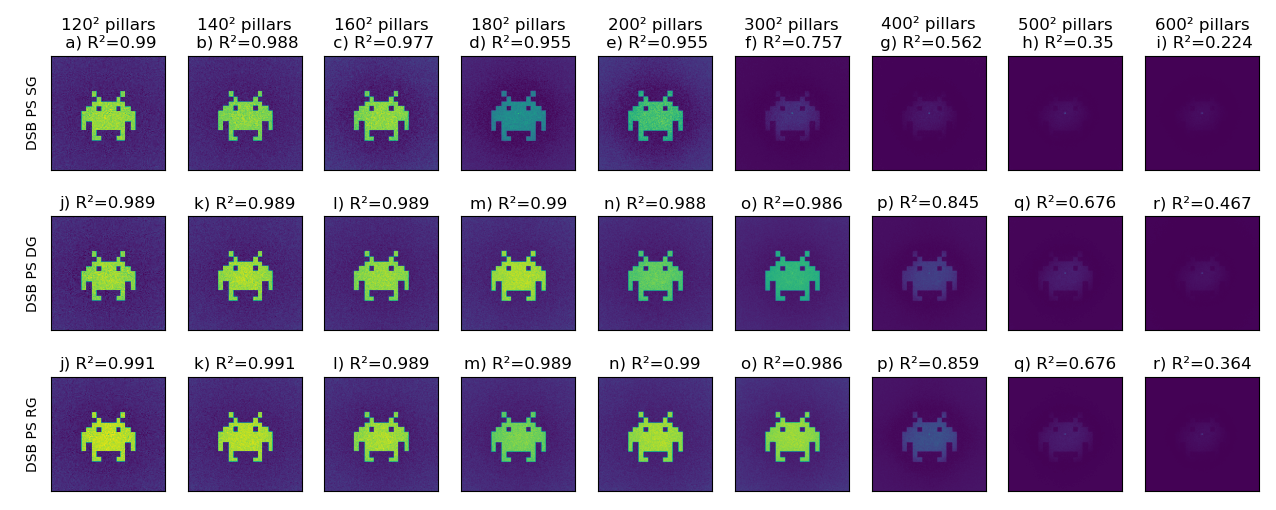}
		\caption{Far field results obtained from FDTD simulations of space invader design using DSB with posterior sampling SG, DG and RG for metasurfaces of increasing size, ranging from $120^2$ to $600^2$ pillars.}
		\label{fig:giga_scaling_SB_space_invader}
	\end{figure}
	
	\begin{figure}[H]
		\hspace{-4cm}
		\includegraphics[scale=0.45]{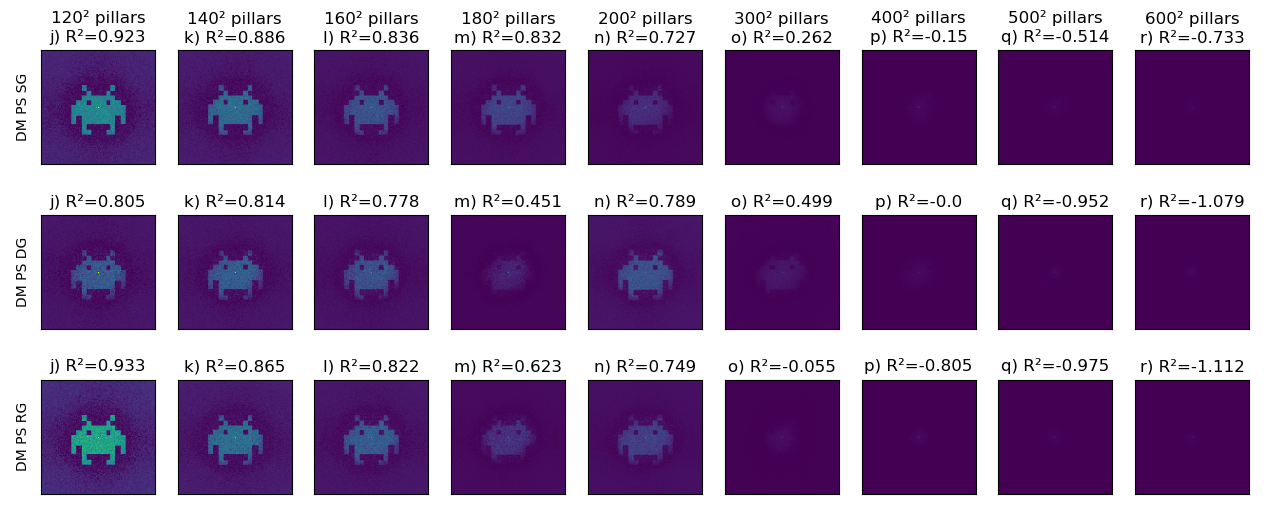}
		\caption{Far field results obtained from FDTD simulations of space invader design using DM with posterior sampling SG, DG and RG for metasurfaces of increasing size, ranging from $120^2$ to $600^2$ pillars.}
		\label{fig:giga_scaling_DM_space_invader}
	\end{figure}
	
	\begin{figure}[H]
		\hspace{-4cm}
		\includegraphics[scale=0.45]{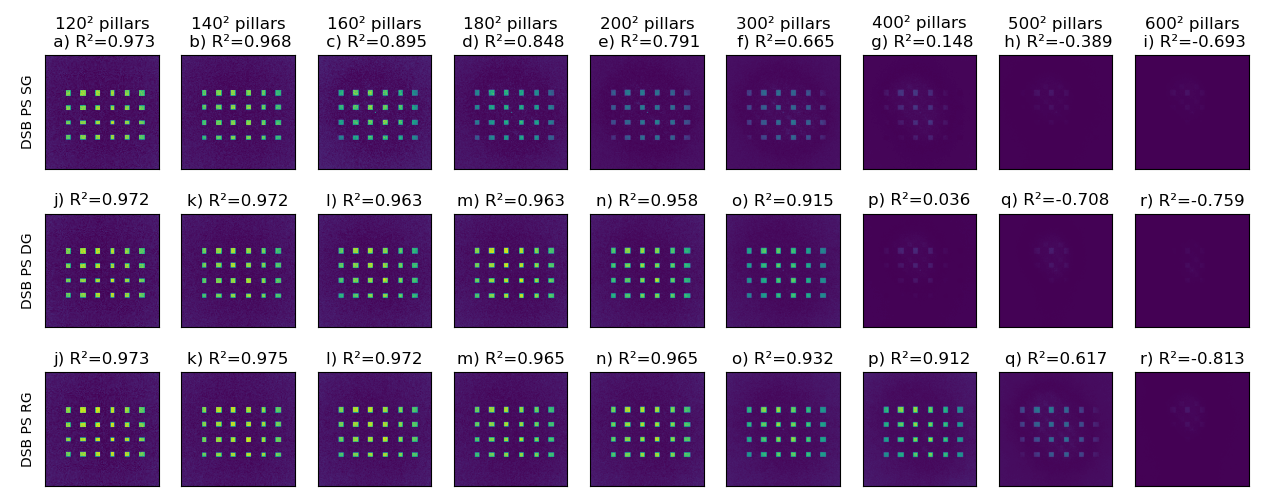}
		\caption{Far field results obtained from FDTD simulations of dots design using DSB with posterior sampling SG, DG and RG for metasurfaces of increasing size, ranging from $120^2$ to $600^2$ pillars.}
		\label{fig:giga_scaling_SB_dots}
	\end{figure}
	
	\begin{figure}[H]
		\hspace{-4cm}
		\includegraphics[scale=0.45]{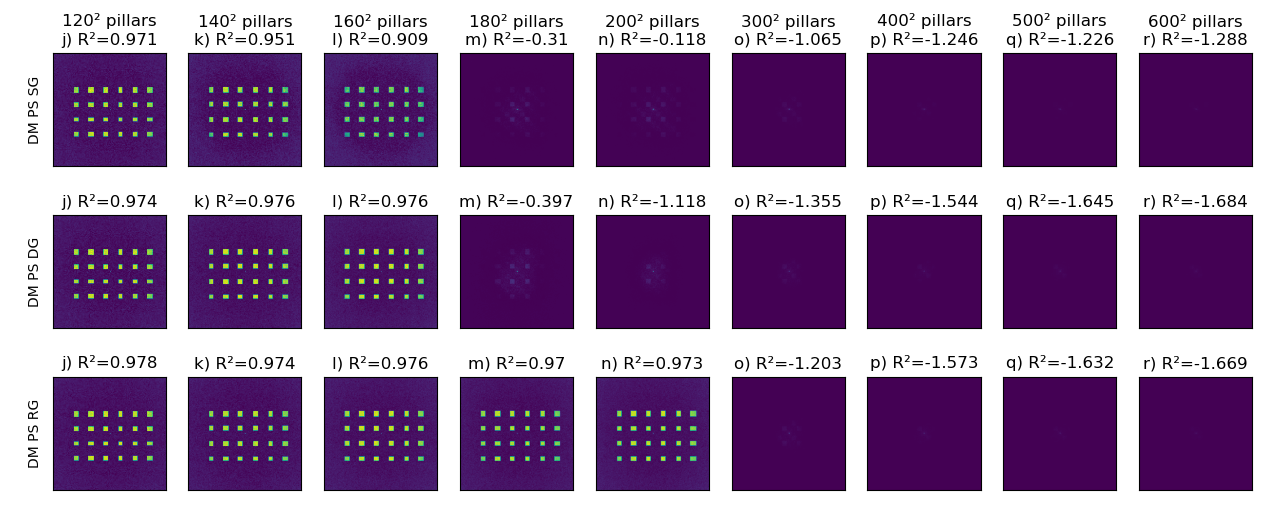}
		\caption{Far field results obtained from FDTD simulations of dots design using DM with posterior sampling SG, DG and RG for metasurfaces of increasing size, ranging from $120^2$ to $600^2$ pillars.}
		\label{fig:giga_scaling_DM_dots}
	\end{figure}
	
	\begin{figure}[H]
		\hspace{-4cm}
		\includegraphics[scale=0.45]{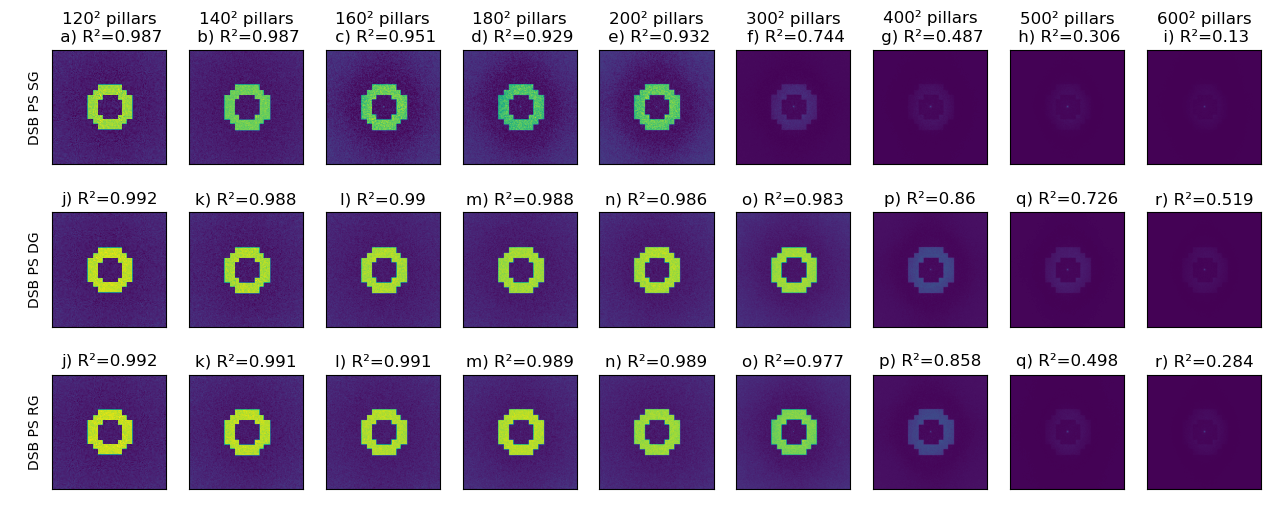}
		\caption{Far field results obtained from FDTD simulations of ring design using DSB with posterior sampling SG, DG and RG for metasurfaces of increasing size, ranging from $120^2$ to $600^2$ pillars.}
		\label{fig:giga_scaling_SB_ring}
	\end{figure}
	
	\begin{figure}[H]
		\hspace{-4cm}
		\includegraphics[scale=0.45]{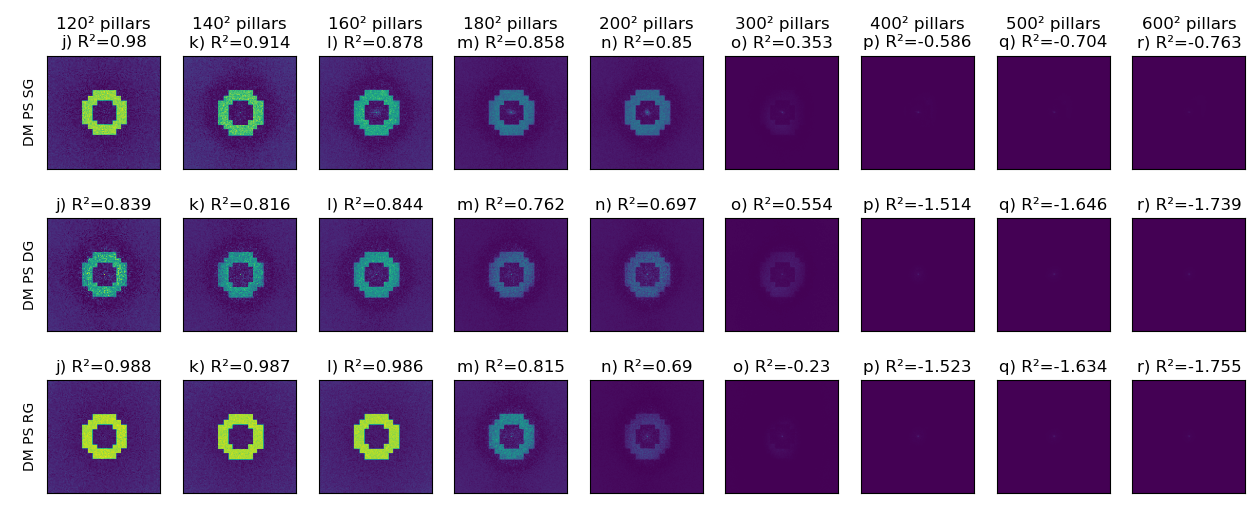}
		\caption{Far field results obtained from FDTD simulations of ring design using DSB with posterior sampling SG, DG and RG for metasurfaces of increasing size, ranging from $120^2$ to $600^2$ pillars.}
		\label{fig:giga_scaling_DM_ring}
	\end{figure}
	
\end{document}